\documentclass[journal]{IEEEtran}
\ifCLASSINFOpdf
\else
\fi
\usepackage{graphicx}
\usepackage{subfig}
\usepackage{amsmath,amssymb}
\usepackage{booktabs,longtable,tabularx,array}
\usepackage{pdflscape}
\usepackage{enumitem}
\usepackage{microtype}
\usepackage{placeins}
\usepackage{multicol}
\usepackage{float}
\newcolumntype{Y}{>{\raggedright\arraybackslash}X}
\newcolumntype{L}[1]{>{\raggedright\arraybackslash}p{#1}}
\begin{document}
\title{Active Hemispherical Metasurface Transmitarray Antenna for Wide-Angle 3D Beam Steering and Target Tracking}
\author{Somayeh~Komeylian,~\IEEEmembership{Member,~IEEE}
and~Christopher~Paolini,~\IEEEmembership{Member,~IEEE}
\thanks{S. Komeylian is with the Department of Electrical and Computer Engineering, 
University of California San Diego, La Jolla, CA 92093 USA, and the Department of Electrical and Computer Engineering, 
San Diego State University, San Diego, CA 92182 USA (e-mail: skomeylian@ucsd.edu).}
\thanks{C. Paolini is with the Department of Electrical and Computer Engineering, 
San Diego State University, San Diego, CA 92182 USA 
(e-mail: paolini@engineering.sdsu.edu).}}
\markboth{}%
{Shell \makeLowercase{\textit{et al.}}: Bare Demo of IEEEtran.cls for IEEE Journals}
\maketitle
\begin{abstract}
A stacked multilayer hemispherical graphene-based transmitarray antenna operating at 250 GHz is developed to achieve wide-angle 3D beam steering.
This work presents the first fully automated CST Studio Suite macro for constructing the multilayer hemispherical geometry and configuring 121 independently tunable graphene sectors.

Numerical results confirm that the integration of voltage-controlled graphene sectors with FCC gold patches enables wide-angle beam steering, achieving an elevation scanning range of $-78^{\circ}$ to $78^{\circ}$ in $\theta$ and full $360^{\circ}$ azimuthal coverage while maintaining stable radiation characteristics and broadband impedance matching ($\left|S_{11}\right| < -10\,\mathrm{dB}$) with a relative bandwidth of 40\%. 
The antenna achieves a directivity range of 11.38--14.72~dBi and an SLL range of $-6.8$ to $-9.8$~dB across all investigated steering directions. 
It also maintains high antenna efficiencies from 71\% to 80\% with a 5\% degradation due to imperfections across the entire scanning range, making it a promising candidate in practical moving-target-tracking applications.
\end{abstract}
\begin{IEEEkeywords}
Graphene surface impedance, metasurfaces, transmitarray antennas, beam steering and reconfigurable antennas.
\end{IEEEkeywords}
\IEEEpeerreviewmaketitle
\section{Introduction}
\IEEEPARstart{A}{ctive} reconfigurable beam-steering architectures have been extensively investigated by several research groups, enabling dynamic control of electromagnetic wavefronts through tunable metasurface elements, Tables~\ref{tab:beam_steering_comparison_1} and ~\ref{tab:beam_steering_comparison}. 
During the early development of graphene-based tunability concepts between 2015 and 2018, several research groups~\cite{balci2018graphene,chen2018graphene}, employed reflectors to demonstrate the feasibility of graphene-enabled electromagnetic reconfigurability.
In this sense, when a plane wave impinges on a surface, the resulting scattered electromagnetic fields are determined by solving Maxwell's equations subject to the appropriate boundary conditions~\cite{Lindell2017,Sandeep2017}. 
For a non-uniform surface, the assumptions underlying Snell's law are not satisfied, as the reflected field is generated by a spatially varying impedance distribution rather than a homogeneous planar interface~\cite{DiazRubio2021}. 
Consequently, the electromagnetic response should be determined using full-wave scattering analysis~\cite{Vahabzadeh2018,Dehmollaian2019,Sandeep2017}; however, such approaches are computationally expensive for complex multilayer and electrically large reflectors. 

Alternatively, when the surface is modeled as a two-dimensional \textbf{periodic} structure, the complexity of the electromagnetic scattering problem is significantly reduced, enabling efficient analysis of the unit-cell response under periodic boundary conditions. 
By exploiting the Floquet--Bloch theorem~\cite{holloway2005frequency} the governing wave equation within the periodic medium is transformed into a closed-form secular equation.
This substantial reduction in computational complexity has established periodic reflectors as an effective framework for analyzing and designing a wide range of electromagnetic systems~\cite{Pozar2012,Boccia2010,Karnati2013}, enabling efficient extraction of the reflection coefficient S$_{11}$ without the need for full-space brute-force scattering analysis.
As an emerging class of tunable reconfigurable reflectors, graphene-based reconfigurable intelligent surfaces (RISs)~\cite{dash2022active,zhang2023risissues}, have attracted significant attention as a promising technology for future 6G networks due to their tunability, compact implementation, particularly because of the aforementioned simplicity in modeling its electromagnetic response through periodic unit-cell analysis.

However, practical RIS deployment architectures, channel models, and integration into cellular systems remain active research areas, since 6G standardization is still in progress and RIS has not yet been adopted as a standardized component in 3GPP specifications~\cite{3gpp6g2025,zhang2023risissues}.
RIS technology has attracted substantial attention as a promising solution for scenarios where direct antenna coverage is compromised by severe blockages, challenging propagation environments, or the prohibitive cost of deploying additional network infrastructure~\cite{wu2020intelligent,basar2019wireless,di2020smart}.
In particular, graphene-based RIS solutions offer a viable approach for mitigating non-line-of-sight (NLoS) communication limitations caused by dense obstacles while providing a compact and potentially cost-effective alternative to conventional base-station or relay deployment in challenging environments.

However, conventional RIS configurations inherently rely on the incident field distribution, causing the reflected wavefront to remain coupled to the spatial amplitude and phase variations of the incoming wave across the aperture.
This inherent coupling restricts independent wavefront manipulation, particularly under non-uniform illumination, where spatial variations in the incident field can degrade beamforming performance  \cite{diRenzo2020smart,huang2019reconfigurable,jiang2019generalized}.

\begin{figure}[htbp]
    \centering    \includegraphics[width=1\columnwidth,height=5cm,keepaspectratio]{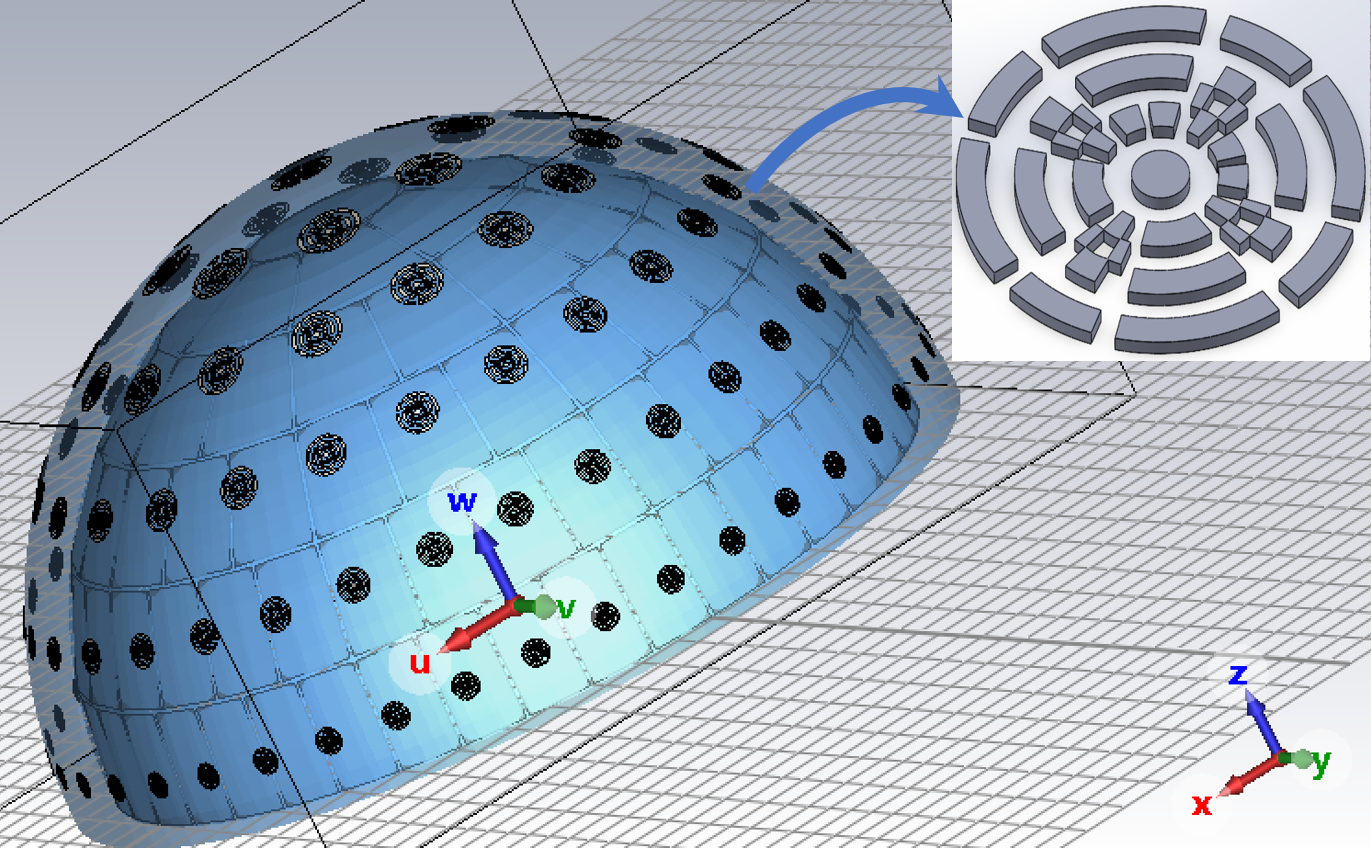}
    \caption{The hemispherical transmitarray antenna comprises 121 FCC elements arranged across the curved aperture, including a central apex element surrounded by five concentric rings containing 8, 16, 24, 32, and 40 patches, respectively. 
    A scaling factor of 0.8 is applied between successive rings to gradually scale the patch dimensions across the aperture. The proposed stacked multilayer transmitarray antenna is composed of a SiO$_2$ substrate with a thickness of $10 \times 10^{-6}$ $m$ and $\varepsilon_{r, SiO_2}=3.9$, followed by an hBN dielectric layer with a thickness of $0.25 \times 10^{-6}$ $m$, graphene sectors, another hBN dielectric layer with a thickness of $0.25 \times 10^{-6}$ $m$ and $\varepsilon_{hBN}=4.4$, and a top layer of FCC patches fabricated from gold (Au) and exposed to free space. The FCC patch geometry and its fractal topology are illustrated in the inset for further clarification ~\cite{komeylian2025high,komeylian2026graphenebasedhemisphericaltransmitarrayantenna}.}
    \label{fig:transmitarray_antenna}
\end{figure}

Unlike conventional reflective RIS architectures, which are mainly designed to manipulate wireless propagation channels and enhance coverage through reflected wave control rather than direct beam synthesis, our proposed hemispherical transmitarray antenna, Fig.~\ref{fig:transmitarray_antenna}, achieves agile electronic beam steering with robust real-time tracking capability in dynamic environments \cite{diRenzo2020smart,wu2020intelligent,zhang2023risissues}.

To overcome these limitations, our proposed hemispherical antenna facilitates full radiation pattern tailoring through voltage-controlled graphene-driven beam steering~\cite{dmitriev2023multifunctional,yadav2025surface}.
In contrast to conventional RIS architectures, our transmitarray antenna further provides a control over the elevation, azimuth, steering angles, and the HPBW. 
Specifically, in the proposed transmitarray antenna, the circular feed horn and hemispherical lens determine the HPBW, whereas the bias voltages applied to the graphene sectors independently regulate the beam-steering direction.

Dynamic radiation-pattern reconfiguration is essential for tracking moving targets in dynamic environments. 
By integrating real-time sensing, trajectory prediction, adaptive positioning, and closed-loop control, a robust tracking system can continuously steer the antenna beam to maintain a high-gain communication link along the target trajectory.

In the proposed architecture, the bias voltages applied to the graphene sectors can be potentially generated by a high-speed FPGA-based control system, enabling real-time electronic beam steering through fast spatial reconfiguration and agile beam deflection.
Consequently, the hemispherical transmitarray antenna can continuously align its main beam with moving targets while preserving high directivity and reliable communication links.

The necessity of dynamically reconfiguring an antenna radiation pattern for moving-target tracking stems primarily from three core operational requirements: 
\textbf{(1) beam steering or directional tracking}: our proposed hemispherical transmitarray antenna dynamically aligns its main radiation beam with the target's time-varying direction of arrival to sustain a stable, high-gain communication link. 
This rapid beam-steering capability is achieved by dynamically controlling the bias voltages applied to the graphene sectors, thereby enabling adaptive wavefront synthesis through precise modulation of the aperture phase distribution, 
\textbf{(2) beamwidth adaptation or HPBW control}: the hemispherical lens and feedhorn enable adaptive HPBW control, providing a broad beam for efficient initial target acquisition and a narrow, high-gain beam for lock-on tracking to maximize the received signal strength and signal-to-noise ratio (SNR),
and 
\textbf{(3) sidelobe and interference suppression}: dynamic radiation-pattern reconfiguration allows the hemispherical transmitarray antenna to actively suppress sidelobe levels and steer nulls toward multipath interference or intentional jamming sources. This adaptive spatial filtering significantly improves communication robustness and target-tracking reliability.

Moreover, in contrast to available reflective RISs~\cite{huang2019reconfigurable,wu2019intelligent,di2020smart}, the proposed antenna functions as a transmissive programmable metasurface lens antenna, where the transmitted wavefront is dynamically controlled to achieve electronic beam steering.

The incident field generated by the feedhorn propagates through the hemispherical multilayer structure and interacts with the tunable graphene-based unit cells and metallic resonators, where controlled phase modulation enables adaptive wavefront synthesis. As a result, the proposed transmitarray antenna produces electronically reconfigurable far-field radiation patterns for agile beam steering.

Another significant advantage of our proposed hemispherical transmitarray antenna over conventional planar transmitarray antennas and reflectors is the intrinsic alignment between the incident wave vector and the local surface normal, which substantially reduces the feed-induced phase compensation required by individual unit cells.
Considering the general monochromatic phase term 
$e^{-j\mathbf{K}\cdot\mathbf{R}}$, the propagation characteristics of an electromagnetic wave in planar and lens-based configurations can be described by the spatial relationship between the propagation vector $\mathbf{K}$ and the position vector $\mathbf{R}$.
Typically, a horn antenna employed as a feed source for either planar~\cite{nayeri2014three,rahmati2017wideband} or curved~\cite{lester1984shaped,ruze1952antenna} configurations illuminates their corresponding apertures with spherical wavefronts, whose spatial propagation is characterized by the wave vector $\mathbf{K}$.

When spherical wavefronts illuminate a planar configuration~\cite{nayeri2014three,pozar1997reflectarray}, the unit cells distributed across the aperture experience nonuniform incident phases due to their different path lengths from the feed source. 
This feed-induced phase variation appears as an additional phase term in the array factor, requiring accurate compensation during aperture phase synthesis to achieve the desired radiation pattern.
In other words, the projection of the propagation vector $\mathbf{K}$ onto the position vector $\mathbf{R}$ of each unit cell results in an additional spatial phase contribution, which should be accounted for in the array factor formulation. 
In contrast, our proposed hemispherical lens configuration places the feedhorn phase center at the center of curvature, ensuring that the propagation vector of $\mathbf{K}$ is approximately aligned with the radius vector $\mathbf{R}$ at every point across the lens aperture. Because both vectors are perpendicular to the local surface, the FCC unit cells experience nearly uniform incident phases, thereby inherently minimizing feed-induced spatial phase variations across the aperture.
As a result, the feed-induced spatial phase variation that arises in planar configurations~\cite{nayeri2014three,rahmati2017wideband,pozar1997reflectarray}, is substantially mitigated in our proposed hemispherical antenna, owing to the intrinsic alignment between the incident propagation vector and the local surface normal across the curved aperture.
Consequently, the intrinsic suppression of feed-induced spatial phase variations in our proposed transmitarray antenna reduces the complexity of aperture phase synthesis, improves aperture efficiency and beamforming accuracy, and enables more efficient voltage-controlled beam steering than conventional planar transmitarray architectures.
\begin{table}[!t]
\caption{Comparison of reflectarray beam-steering mechanisms based on S$_{11}$ phase response and scanning range.}
\label{tab:beam_steering_comparison_1}
\centering
\footnotesize
\renewcommand{\arraystretch}{1.05}

\begin{tabular}{p{0.65cm} p{0.65cm} p{1.8cm} p{3.2cm} p{1.2cm}}
\hline
\textbf{Ref.} &
\textbf{Freq.} &
\textbf{Reflector Type} &
\textbf{Beam-Reconfiguration Mechanism} &
\textbf{Beam Scanning Range} \\
\hline

\cite{qureshi2024reflectarray} &
125 GHz &
Graphene reflectarray with an aperture diameter of $15\lambda$&
Graphene chemical-potential tuning through DC biasing dynamically controls the reflection phase of resonant graphene elements. &
$\pm45^\circ$ \\

\cite{chauhan2025sub} &
0.27 THz &
Graphene RIS with an
aperture size of $13.6\lambda\times13.6\lambda$ &
Chemical-potential-controlled graphene conductivity provides discrete reflection phase states for programmable beam steering. &
$0^\circ$--$40^\circ$ \\

\cite{esquius2014sinusoidally} &
1 THz &
Graphene leaky-wave antenna. The aperture dimensions are not explicitly specified&
DC bias tuning of graphene conductivity modifies the surface reactance and controls the beam direction. &
$\approx\pm30^\circ$ \\

\cite{rubio2022mechanically} &
14 GHz &
Mechanical reflectarray with an aperture size of $12\lambda \times 12\lambda$ &
Flexible patch elements on a plastic substrate are mechanically deformed to modify the phase distribution. &
$\pm21^\circ$\\

\cite{hosseininejad2019digital} &
$\sim$1 THz &
Graphene digital metasurface. The aperture dimensions are not explicitly specified &
Chemical-potential tuning creates programmable reflection phase gradients for beam steering. &
$\approx\pm60^\circ$ \\

\textbf{This WORK} &
\textbf{250 GHz} &
\textbf{Graphene metasurface transmitarray with an aperture diameter of $\approx 4.17\lambda$} &
\textbf{Voltage-controlled graphene chemical-potential tuning modifies local surface impedance and transmission phase of FCC resonant unit cells.} &
\textbf{Elevation angle: $\pm78^\circ$, and full azimuthal coverage: $360^\circ$}\\

\hline
\end{tabular}
\end{table}

\begin{table}[!t]
\caption{Comparison of transmissive beam-steering antennas based on S$_{21}$ phase response and scanning capability.}
\label{tab:beam_steering_comparison}
\centering
\footnotesize
\renewcommand{\arraystretch}{1.05}

\begin{tabular}{p{0.65cm} p{0.65cm} p{1.7cm} p{3.3cm} p{1.2cm}}
\hline
\textbf{Ref.} &
\textbf{Freq.} &
\textbf{Antenna Type} &
\textbf{Beam-Reconfiguration Mechanism} &
\textbf{Beam Scanning Range} \\
\hline

\cite{li2019wide} &
5.75 GHz &
Conformal metasurface lens antenna with an aperture size of $\approx 7.9\lambda \times 4.1\lambda$ &
DC bias voltage tuning of varactor-loaded metasurface elements enables programmable transmission-phase control. Multiple feed sources further extend the steering range. &
$\pm60^\circ$
\\[1.5ex]

\cite{madannejad2025passive} &
610--685 GHz &
Planar Fresnel lens antenna with an aperture size of $\approx 13.17\lambda \times 13.17\lambda$ &
Passive frequency-dependent beam steering using a Fresnel zone planar lens combined with a graded-index silicon perforated disk. &
$-30^\circ$ to $-14^\circ$
\\[1.5ex]

\cite{omam2025wideband} &
24--40 GHz &
Flat dielectric lens antenna (FDL antenna) with a normalized aperture size of $\frac{D}{\lambda_0}=11.1$, where $D\approx104.1~\mathrm{mm}$&
Passive multi-feed beam steering using a nine-element waveguide feed network positioned along the focal arc of a flat dielectric lens. &
$\pm33^\circ$

\\[1.5ex]

\cite{alonso2024transmit} &
450--650 GHz &
Transmit lens antenna array with an aperture size of $\approx 28.3\lambda$ &
Mechanical translation of a silicon transmit lens array using a piezo actuator to modify the relative phase distribution. &
$\pm25^\circ$
\\[1.5ex]

\cite{keshavarz2026ultra} &
2.1 GHz &
Microstrip Arrayed Lines Lens (MALL) with an aperture size of $0.48\lambda \times 0.52\lambda$ &
Passive beam steering using an interdigital-loaded Microstrip Arrayed Lines Lens (MALL) with controlled transmission-line phase delay. &
$\pm60^\circ$
\\[1.5ex]

\cite{kitayama2026transmissive} &
115 GHz &
Transmissive liquid-crystal metasurface with an aperture size of $\approx 27.25\lambda \times 27.25\lambda$ &
Electrical control of liquid-crystal molecular orientation dynamically tunes the phase/amplitude response of subwavelength unit cells. &
$\pm30^\circ$
\\[1.5ex]
\textbf{This WORK} &
\textbf{250 GHz} &
\textbf{Hemispherical metasurface transmitarray antenna with an aperture diameter of $\approx 4.17\lambda$} &
\textbf{Voltage-controlled graphene chemical potential tuning modifies the local surface impedance and transmission phase of FCC resonant unit cells.} &
\textbf{Elevation angle: $\pm78^\circ$, and full azimuthal coverage: $360^\circ$}
\\
\hline
\end{tabular}
\end{table}
To further highlight the functionality and advantages of the proposed transmitarray antenna, the following discussion presents the differences between reflective metasurfaces and transmitarrays.
In active beam-steering systems, reflective metasurfaces, such as RISs, can achieve wide angular scanning ranges through programmable control of the reflected wavefront; however, their performance remains inherently coupled to the incident field distribution.
This behavior results from the S$_{11}$-based beamforming mechanism, in which the incident electromagnetic wave interacts primarily with the surface impedance, eliminating the additional phase variations associated with propagation through multilayer transmission structures.
Consequently, the electromagnetic interactions among the surface elements remain relatively weak, allowing larger phase gradients to be achieved and, consequently, enabling wider beam-scanning angles.
Nevertheless, reflective architectures are primarily suited for laboratory demonstrations, proof-of-concept implementations, or applications requiring the tracking of a limited number of targets.

In contrast, transmissive metasurface antennas are governed by the transmission coefficient S$_{21}$, requiring the incident electromagnetic wave to propagate through multiple dielectric and conductive layers, where inter-layer coupling and propagation effects collectively determine the transmitted phase and amplitude response.
These internal electromagnetic interactions, including multiple reflections, transmission losses, impedance mismatches, and interlayer coupling, severely restrict the achievable transmission phase range, amplitude response, and overall beamforming performance.
Consequently, owing to the additional propagation and multilayer electromagnetic interactions, transmissive metasurface antennas generally achieve narrower beam-scanning ranges than reflective metasurfaces, despite offering greater flexibility for wavefront synthesis and radiation control.
Despite this limitation, transmitarray antennas offer distinct advantages for deployment in wireless communication systems.

They generally provide higher antenna gain, higher antenna efficiency, more efficient utilization of the incident electromagnetic power, and full-space transmission, making them particularly suitable for high-capacity wireless communication systems.
These characteristics render transmitarray antennas ideally suited for emerging 6G wireless networks and broadband satellite constellations, such as Starlink, where high antenna gain, efficient power utilization, and reliable communication links are often prioritized over maximizing the beam-scanning range.
Although reflective metasurfaces achieve wider scanning angles, transmitarrays offer a more viable real-world trade-off between beam-steering flexibility, gain, and overall antenna efficiency.
In this work, the hemispherical lens of the proposed transmitarray antenna introduces an additional degree of freedom for wavefront control, thereby achieving a wider beam-steering range than conventional planar reflectarrays, Table~\ref{tab:beam_steering_comparison_1}, and transmitarrays, Table~\ref{tab:beam_steering_comparison}.
The remainder of this paper is organized as follows.  
Section II presents the design objectives and physical architecture of the proposed hemispherical graphene-based transmitarray antenna. 
Section III is devoted to the CST Studio Suite visual basic for applications (VBA) macro developed for automated modeling and construction of the multilayer hemispherical transmitarray. 
Section IV provides the performance evaluation and discussion, including beam-steering characteristics, radiation performance, antenna efficiency, and impedance bandwidth. 
Finally, Section V presents a brief conclusion of this work.
\section{Design Specifications and Physical Architecture}
Our proposed antenna is a feedhorn-excited programmable hemispherical transmissive metasurface lens operating at 250 GHz, Fig.~\ref{fig:transmitarray_antenna}. 
The lens modifies the aperture field distribution to achieve an enhancement in its radiation pattern control, a reduction in its sidelobe levels (SLLs) by suppressing the radiation pattern away from the desired main beam and an increased angular discrimination between the main lobe and sidelobes.
Our proposed hemispherical lens is realized as a conformal stacked multilayer architecture consisting of a SiO$_2$ dielectric supporting substrate, voltage-controlled graphene sectors, ultra-thin hexagonal boron nitride (hBN) dielectric isolation layers, and fractal concentric circular (FCC) gold resonators~\cite{komeylian2025high}, serving as frequency-selective surface (FSS) elements, as illustrated in Fig.~\ref{fig:transmitarray_antenna}.
The electromagnetic functionalities and contributions of the individual constituent layers are outlined as follows:

\textbf{Conformal FCC Gold Resonators}: 
The FCC elements function as localized resonant FSS elements electromagnetically coupled to the underlying graphene sectors.
The scaling factor $\alpha$ was optimized through a parametric sweep to establish an appropriate trade-off between geometric miniaturization and electromagnetic performance. 
Extremely small scaling factors, such as $\alpha=0.1$, were excluded because the pronounced dimensional reduction between successive fractal levels imposes geometric constraints that limit the ability of the fifth-ring elements to maintain the required transmission amplitude and phase responses. 
In contrast, $\alpha=1$ was also excluded since the dimensional scaling between adjacent rings is essential for achieving the desired bandwidth enhancement and fully exploiting the advantages of the fractal geometry. Accordingly, the range $0.3\leq\alpha\leq0.9$ was selected for the parametric optimization. A systematic parametric sweep of the candidate scaling factors showed that $\alpha=0.8$ provides the most favorable balance between fractal miniaturization and electromagnetic performance for $a=8$. 
The resulting geometry and ring-dependent dimensions are summarized in Table I of ~\cite{komeylian2025high}. Further design principles and electromagnetic advantages of the FCC elements, particularly their contribution to bandwidth enhancement and resonant response engineering, are discussed in detail in ~\cite{komeylian2025high,komeylian2026graphenebasedhemisphericaltransmitarrayantenna}.

\textbf{Graphene Sectors}: 
The graphene sectors provide electrically tunable surface impedance via voltage-controlled modulation of the chemical potential.
They function as voltage-controlled tunable impedance surfaces, where the applied bias voltages independently regulate the chemical potentials of the graphene sectors, thereby modifying their complex surface conductivities and enabling dynamic control of the electromagnetic response of each unit cell.
This electrical tunability provides dynamic control over local transmission amplitude and phase responses, enabling programmable beam steering.

\textbf{hBN Dielectric Layers}:
The proposed stacked multilayer architecture employs two hBN dielectric layers, designated as hBN$_{1}$ and hBN$_{2}$.
The hBN$_1$ layer is positioned between the SiO$_2$ substrate and the graphene sectors, serving as an electrical isolation layer and dielectric spacer. 
During the fabrication process, microchannels can be implemented within hBN$_1$ to provide independent voltage biasing of individual graphene sectors without altering the SiO$_2$ supporting layer.
The hBN$_2$ layer, positioned between the graphene sectors and FCC gold resonators, prevents direct electrical contact and short circuits while preserving the required electromagnetic coupling for tunable unit-cell operation.

\textbf{SiO$_2$ Supporting Layer}:
The hemispherical dielectric substrate of SiO$_2$ provides mechanical support for the conformal metasurface. 
Moreover, this hemispherical profile introduces controlled dielectric loading and enforces the physical curvature necessary for efficient radial-to-planar wavefront transformation.

\textbf{Circular Feedhorn Antenna}:
The excitation feedhorn, with an aperture diameter of 5 $mm$, is positioned at the geometric center of the hemispherical aperture so that the incident electromagnetic wave illuminates the conformal metasurface predominantly along the local radial direction. 
This geometrically symmetric placement minimizes aperture phase nonuniformity, equalizes the propagation path lengths to the FCC resonant patches, and maximizes the electromagnetic coupling efficiency between the feed source and the hemispherical transmitarray, thereby improving wavefront uniformity and transmission performance.

Unlike conventional RISs, our proposed antenna, Fig.~\ref{fig:transmitarray_antenna}, operates as a transmissive programmable metasurface lens. 
As the electromagnetic wave radiated by the feedhorn propagates through the stacked multilayer hemispherical aperture, it interacts with the electrically tunable graphene sectors and FCC resonators to modulate the output field.
By dynamically adjusting the local transmission profile of individual unit cells, the transmitted wavefront is reshaped to synthesize the desired far-field radiation pattern with electronically reconfigurable beam-steering capabilities.
\section{Analytical Framework for the Voltage-Controlled Surface Impedance and Equivalent RLC Circuit of Stacked Multilayer Unit Cells}
The electromagnetic conductivity of graphene is described by the Kubo formalism, which consists of intraband and interband contributions~\cite{nayyeri2013wideband,cano2025accurate}. 
The intraband conductivity originates from the Drude-like response of graphene charge carriers and governs the electromagnetic behavior of graphene at microwave and terahertz frequencies. 
This contribution enables electrically tunable conductivity through modulation of the graphene chemical potential, thereby providing dynamic control over the surface impedance and electromagnetic response.
In contrast, the interband conductivity is associated with electronic transitions between the valence and conduction bands and primarily contributes at higher frequencies, such as the infrared and optical regimes, where interband excitation becomes dominant.

Therefore, the intraband term alone accurately approximates graphene conductivity in most terahertz transmitarray applications, while the interband contribution is negligible for most terahertz transmitarray applications.  

For illustrative purposes, if the operating frequency were shifted from 250 GHz to the infrared or optical regimes, dielectric excitation of the hBN layers would become significant.
In this sense, inserting an hBN layer between FCC metal patches and graphene introduces resonator regions, providing additional phase variations across the hemispherical curvature of 
In other words, varying the bias voltages applied to the graphene sectors modifies not only does the graphene surface conductivity but also the electromagnetic coupling with the surrounding multilayer structure.

In other words, not only does a variation in the bias voltages applied to the graphene sectors modify the graphene surface conductivity but it also modifies the electromagnetic coupling with the surrounding multilayer structure.
In particular, the resonant response of the hBN dielectric layer between the metallic FCC patches and graphene introduces additional reactive loading, providing an extra degree of freedom for controlling the transmission phase.
By electrically tuning the resonant interaction between the graphene sectors and the hBN dielectric layer, the transmitted phase can be precisely controlled beyond the graphene conductivity variation alone. 
This additional reactive tuning mechanism provides an extra degree of freedom for phase engineering and enables the realization of a smooth phase gradient over the metasurface aperture.
It is worth emphasizing that the coupling between the resonant modes of the hBN layer and the intraband and interband electronic transitions in graphene becomes significant primarily at higher frequencies, particularly in the infrared and optical regimes, where the photon energy is sufficient to excite these electronic transitions.
However, at 250 GHz, the photon energy remains below the interband transition threshold, and the graphene response is predominantly governed by the tunable intraband Drude conductivity. Hence, graphene conductivity is dominated by the intraband Drude contribution, while the interband effect is negligible. 
Because hBN dielectric excitation occurs only in the infrared and optical regime, the hBN layer cannot resonantly couple to the graphene layer and therefore does not provide an additional degree of freedom for the phase tuning at 250 GHz. 
Consequently, the response of the stacked multilayer unit cells is governed primarily by graphene, where modulating the graphene chemical potential alters its intraband conductivity to enable dynamic transmission phase control.
Because the hBN layer behaves as a conventional dielectric rather than a resonant medium at 250 GHz, it does not introduce additional resonant phase modulation into the response of the stacked multilayer unit cells. 
Consequently, beam-steering and beam-squint mitigation mechanisms associated with hBN dielectric resonances, which become significant only at higher frequencies, are limited in the present sub-terahertz or low-terahertz regime and cannot be effectively utilized.
At 250 GHz, the tunable electromagnetic response is therefore dominated by the voltage-controlled graphene conductivity and its associated surface impedance variation.

\begin{figure}[htbp]
    \centering
    \includegraphics[width=1\linewidth]{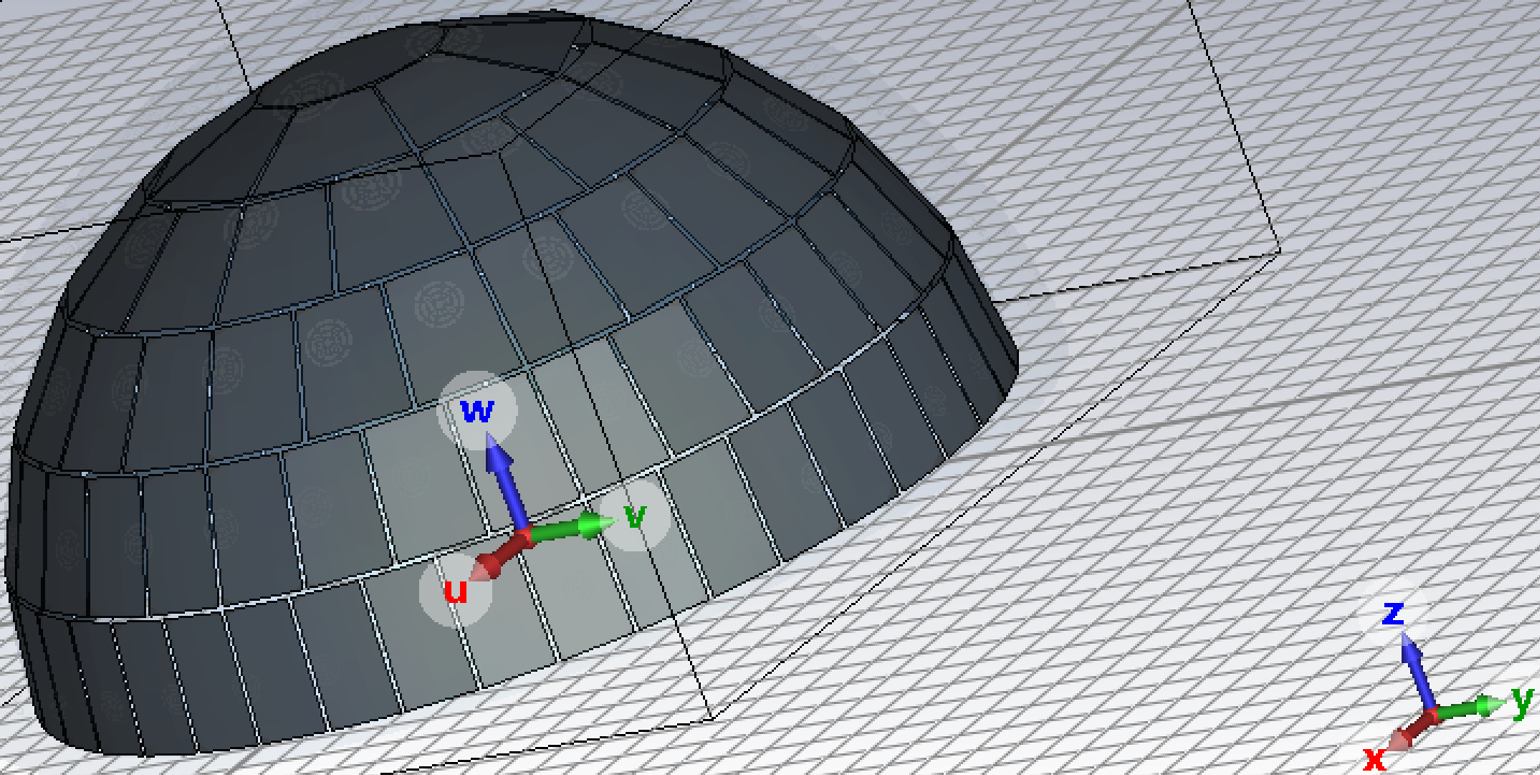}
    \caption{The hemispherical transmitarray comprises 121 independently tunable graphene sectors distributed across the curved aperture, where each unit cell is modeled as a locally planar trapezoidal sector. 
    Each graphene sector is independently biased through a gate voltage to dynamically regulate its chemical potential, enabling tunable electromagnetic responses over the range of $0.1~\mathrm{eV} \leq \mu_{c,(n,k)} \leq 1~\mathrm{eV}$. All unit cells are characterized by a common carrier relaxation time of  $\tau=0.1$ps and an operating temperature of $T=300K$. The graphene sectors are physically and electrically isolated from neighboring elements, ensuring independent electrostatic bias control and reliable reconfigurable electromagnetic operation~\cite{komeylian2026graphenebasedhemisphericaltransmitarrayantenna}.}
    \label{fig:graphene-sectors}
\end{figure}

\begin{figure}[htbp]
    \centering    \includegraphics[width=1\columnwidth,height=5cm,keepaspectratio]{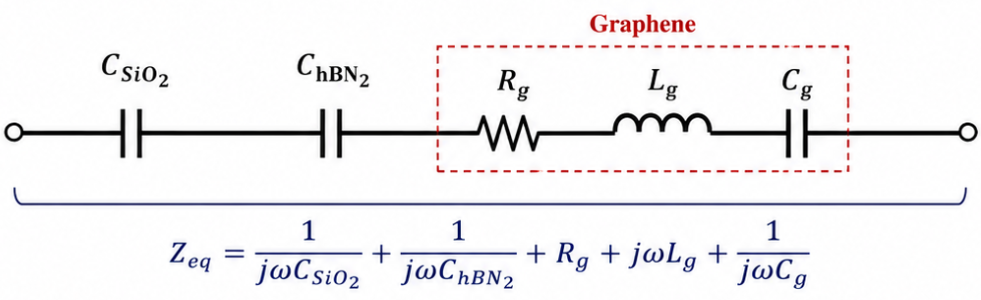}
    \caption{The equivalent circuit model of our proposed hemispherical transmitarray antenna corresponding to Fig.~\ref{fig:transmitarray_antenna}. The model includes the dielectric capacitances of the SiO$_2$ substrate, $C_{\mathrm{SiO_2}}$, and upper hBN layer, $C_{\mathrm{hBN}_2}$, while the graphene sheet impedance is represented by a series $R_g$-$L_g$ branch with an additional gap capacitance $C_g$. The lower hBN layer is omitted because it contributes only a fixed capacitive loading effect due to its invariant dielectric properties~\cite{komeylian2026graphenebasedhemisphericaltransmitarrayantenna}}.
    \label{fig:RLC_Circuit}
\end{figure}

In the following, we briefly review the analytical framework of our transmitarray antenna. The comprehensive theoretical formulations, their implications, and the corresponding formulation derivations are discussed in detail in~\cite{komeylian2026graphenebasedhemisphericaltransmitarrayantenna}.
The beam-steering capability of the proposed transmitarray antenna, Fig.~\ref{fig:graphene-sectors}, originates from the electrically tunable graphene 
surface impedance. 
The graphene surface impedance can be dynamically controlled by the intraband Kubo model, where the chemical potential $\mu_c$ modulates the static conductivity of $\sigma_0(\mu_c)$ to determine the overall complex conductivity of the graphene sheet~\cite{nayyeri2013wideband,komeylian2025high,komeylian2026graphenebasedhemisphericaltransmitarrayantenna},
\begin{equation}
\begin{aligned}
\sigma_{\mathrm{intra}}(\omega,\mu_c,\Gamma,T)
&=
-j\frac{e^2 k_B T}
{\pi\hbar^2(\omega-j2\Gamma)}
\\
&\quad\times
\left[
\frac{\mu_c}{k_B T}
+
2\ln\left(
1+e^{-\mu_c/(k_B T)}
\right)
\right]
\end{aligned}
\label{eq:kubo}
\end{equation}
where $\mu_c$, $\tau$, $T$, $k_B$, $\hbar$ represent the graphene chemical potential, carrier relaxation time, temperature, Boltzmann constant, and reduced Planck constant, respectively. $\Gamma$ is related to $\tau$ by $\Gamma = \frac{1}{2\tau}$. According to the intraband (Drude-like) Kubo conductivity of graphene in Eq.\ref{eq:kubo}, the graphene conductivity varies continuously with the chemical potential, which is controlled by the voltages applied to the graphene sectors. 
The electrostatic relation between the voltage and graphene chemical potential is expressed as~\cite{komeylian2026graphenebasedhemisphericaltransmitarrayantenna}, 
\begin{equation}
\mu_{c,(n,k)}\;(\mathrm{eV})
=
\frac{\hbar v_F}{e}
\sqrt{\frac{\pi C_g}{e}
\left|V_{g,(n,k)}-V_D\right|}
\label{eq:chemical_potential}
\end{equation}
where $v_F$, $C_g$, $V_{g,(n,k)}$, and $V_D$ denote the Fermi velocity of graphene, the gate capacitance per unit area, the applied gate voltage of the $(n,k)$-th unit cell, and the Dirac-point voltage, respectively.
At 250 GHz, considering only the graphene sheet without the surrounding multilayer structure, the equivalent surface impedance is directly determined from the surface conductivity using the intraband Kubo formulation, as follows~\cite{komeylian2026graphenebasedhemisphericaltransmitarrayantenna},
\begin{equation}
Z_s(\omega,\mu_c,\Gamma,T)
\approx
\frac{1}{\sigma_{\mathrm{intra}}(\omega,\mu_c,\Gamma,T)}
\label{eq:Zs_graphene}
\end{equation}
The graphene surface conductivity is governed primarily by the intraband Drude response, while the interband contribution is negligible over the considered sub-THz range. 
In this case, the resistive component originates from finite carrier scattering, whereas the imaginary component represents the kinetic inductance associated with the inertia of charge carriers. 
Since the lateral dimensions and total thickness of each stacked multilayer unit cell are electrically small compared with the operating wavelength,  the multilayer structure including FCC patch layer, hBN dielectric isolation layers, and supporting substrate can be accurately modeled as an equivalent impedance sheet characterized by the following complex surface impedance ~\cite{komeylian2026graphenebasedhemisphericaltransmitarrayantenna}, 
\begin{equation}
\begin{split}
Z_{s,(n,k),\xi}(V_g)
&=
\underbrace{\frac{D_{\xi'}\left(1+\omega^2\tau^2\right)}
{G_{\xi}\,\sigma_{0,(n,k)}}
}_{R_{\xi}(V_g)}
+
j\omega\,
\underbrace{
\frac{\tau D_{\xi'}}
{G_{\xi}\,\sigma_{0,(n,k)}}
}_{L_{\xi}(V_g)}
\\
&\quad
-
\frac{j}{\omega}
\underbrace{
\left(
\frac{1}{C_{\mathrm{geo},(n,k),\xi}}
+
\frac{\pi\hbar^2 v_f^2}
{2e^2D_{\theta}D_{\phi}
\left|\mu_{c,(n,k)}\right|}
\right)
}_{\displaystyle \frac{1}{C_{\mathrm{eff},\xi}(V_g)}}
\end{split}
\label{eq:voltage_controlled_surface_impedance}
\end{equation}
where $\xi \in \{\theta,\phi\}$ represents the polarization index, corresponding to the $\theta$- or $\phi$-polarized electromagnetic field component and $\xi'$ denotes the orthogonal periodicity (e.g., for $R_{\theta},\qquad D_{\xi'} = D_{\phi}$). Here, $D_{\theta}$ and $D_{\phi}$ denote the angular discretization factors associated with the dimensions of sectors along the curved aperture.
The effective capacitance of $C_{\mathrm{eff},\xi}$ consists of two components of (a) geometrical capacitance $C_{\mathrm{geo},(n,k),\xi}$ and (b) quantum capacitance $C_q$.
The conductivity variation changes the graphene resistance $R_{\xi}$, kinetic inductance $L_{\xi}$, and quantum capacitance $C_{q}$.
Independently tuning the impedance of spatially distributed graphene sectors synthesizes the phase profile required to steer transmitted electromagnetic waves toward directions of moving targets.
The multilayer stack surrounding the graphene sectors modifies the local electromagnetic field distribution and current flow, introducing additional equivalent inductive and capacitive loading that influences the resonant patches and the resulting effective surface impedance.
Hence, the complete voltage-controlled impedance model of the stacked multilayer transmitarray unit cell is established by incorporating the electromagnetic loading effects of the graphene sectors and surrounding dielectric and metallic layers, consistent with Fig.~\ref{fig:transmitarray_antenna}, in the following equation ~\cite{komeylian2026graphenebasedhemisphericaltransmitarrayantenna},  
\begin{equation}
Z_{\mathrm{cell}}(V_g)
=
Z_g(V_g)
+
Z_{\mathrm{hBN}_2}
+
Z_{\mathrm{SiO}_2}
+
\frac{1}{j\omega C_g}
\label{eq:cell_impedance}
\end{equation}
Consequently, the complete electromagnetic response of the unit cell can be efficiently represented by an equivalent RLC circuit model in Fig.~\ref{fig:RLC_Circuit}.
Accordingly, the impedance contributions of the SiO$_2$ substrate and upper hBN dielectric layer are given by~\cite{komeylian2026graphenebasedhemisphericaltransmitarrayantenna}, 
\begin{equation}
Z_{\mathrm{SiO}_2}
=
j\,\frac{\eta_0}{\sqrt{\varepsilon_{r,\mathrm{SiO}_2}}}
\tan\!\left(
k_0\sqrt{\varepsilon_{r,\mathrm{SiO}_2}}\,
t_{\mathrm{SiO}_2}
\right)
\end{equation}
\begin{equation}
Z_{\mathrm{hBN}_2}
=
j\,\frac{\eta_0}{\sqrt{\varepsilon_{r,\mathrm{hBN}}}}
\tan\!\left(
k_0\sqrt{\varepsilon_{r,\mathrm{hBN}}}\,
t_{\mathrm{hBN}_2}
\right)
\end{equation}
where $\eta_0$ and $k_0$ refer to the free-space impedance and the free-space wavenumber, respectively. $t_{(SiO_2)}$ and $t_{(hBN_2)}$ denote the thicknesses of the corresponding dielectric layers. 
This electrically controlled conductivity modulates the effective surface impedance of each transmitarray unit cell, which subsequently tunes the transmission coefficient, $S_{21}$, particularly its transmission phase, thereby enabling dynamic beam steering of the transmitted wavefront \cite{yadav2025surface,dmitriev2023multifunctional,dash2022active}, in the following sequence of the tuning mechanism~\cite{komeylian2026graphenebasedhemisphericaltransmitarrayantenna}, 
\begin{equation}   
V_g
\;\rightarrow\;
\mu_c
\;\rightarrow\;
\sigma_g
\;\rightarrow\;
Z_s
\;\rightarrow\;
\angle S_{21}
\;\rightarrow\;
\text{Beam Steering}
\end{equation}

\subsection{Spherical Array Factor and Beam Performance Metrics}In contrast to planar arrays, where the phase distribution can be described by a linear phase gradient along two orthogonal axes, our proposed hemispherical aperture requires a 3D phase formulation due to the nonplanar distribution of the radiating elements. 
Each unit cell introduces a unique phase contribution determined by its spherical position and propagation path. Therefore, the conformal array factor should incorporate the exact 3D coordinates of all elements to accurately predict the far-field radiation characteristics. Accordingly, the conformal array factor is expressed as~\cite{komeylian2026graphenebasedhemisphericaltransmitarrayantenna},
\begin{equation}
\begin{aligned}
AF(\Theta,\Phi;p)
={}&
\underbrace{
W_0 e^{jk_0\hat{\mathbf{s}}\cdot\mathbf{r}_0}
}_{\text{Apex element}}
+
\underbrace{
\sum_{n=1}^{N_r}
\sum_{k=1}^{N_n}
W_{n,k}
e^{jk_0\hat{\mathbf{s}}(\Theta,\Phi)\cdot\mathbf{r}_{(n,k)}}
}_{\text{Concentric ring elements}}
\end{aligned}
\label{eq:AF}
\end{equation}
where $\hat{\mathbf{s}}(\Theta,\Phi)$ and $\mathbf{r}_{(n,k)}$ represent the observation unit vector and the position vector of the $(n,k)$-th unit cell on the hemispherical aperture, respectively.
The complex excitation coefficient of each element is expressed in the following equation,~\cite{komeylian2026graphenebasedhemisphericaltransmitarrayantenna},
\begin{equation}
W_{(n,k)}
=
\Delta S_n
F_{\mathrm{horn}}(\theta_n,\phi_{(n,k)})
F_{\mathrm{cell},(n,k)}(\Theta,\Phi)
S_{21,(n,k)}
\label{eq:weight}
\end{equation}
where $\Delta S_n$, $F_{\mathrm{horn}}(\theta_n,\phi_{(n,k)})$, and $F_{\mathrm{cell},(n,k)}(\Theta,\Phi)$ denote the effective area of the $(n,k)$-th unit cell, the incident field distribution illuminated by the feed horn, and the radiation pattern of the corresponding unit cell, respectively. 
$S_{21,(n,k)}$ incorporates the programmable transmission amplitude and phase of the graphene unit cell. 
During the initial optimization stage, the element pattern may be approximated by $F_{\mathrm{cell},(n,k)}\approx1$, to focus primarily on the aperture illumination and programmable transmission response. 
After incorporating the effects of feedhorn illumination, the conformal hemispherical geometry, and the graphene transmission characteristics, the actual main-beam direction is determined by the direction of maximum radiated power and is expressed as, 
\begin{equation}
(\Theta_{\mathrm{(pk)}},\Phi_{\mathrm{(pk)}})
=
\arg\max_{\Theta,\Phi}
|AF(\Theta,\Phi)|^{2}
\label{eq:peak}
\end{equation}
However, the beam-steering performance cannot be fully characterized by the main-beam direction alone. A comprehensive assessment requires additional beam-quality metrics that evaluate steering accuracy, gain degradation, and sidelobe characteristics. 
The angular steering errors relative to the desired beam direction of $(\Theta_0,\Phi_0)$ are defined as,
\begin{equation}
\delta\Theta
=
\left|
\Theta_{\mathrm{(pk)}}
-
\Theta_0
\right|
\label{eq:dtheta}
\end{equation}
\begin{equation}
\delta\Phi
=
\min_m
\left|
\Phi_{\mathrm{(pk)}}
-
\Phi_0
+
2\pi m
\right|
\label{eq:dphi}
\end{equation}
where Eq.\ref{eq:dphi} consider the periodicity of the azimuth angles. $m$ denotes an integer used to account for the periodicity of the azimuth angle. 
In this work, the typical angular pointing errors remain within $\delta\Theta \leq 1^\circ$ for elevation steering beam and $3^\circ \leq \delta\Phi \leq 5^\circ$ for azimuth steering beam. 

The relative gain degradation, $G_{\mathrm{rel}}$, is another important metric for evaluating beam steering performance. 
It quantifies the scan loss by comparing the maximum magnitude of the array factor at the steered direction with the broadside reference response, as expressed by,
\begin{equation}
G_{\mathrm{rel}}
=
20\log_{10}
\left(
\frac{|AF(\Theta_{\mathrm{(pk)}},\Phi_{\mathrm{(pk)}})|}
{|AF^{ref}_{\mathrm{broadside}}|}
\right)
\label{eq:grel}
\end{equation}
where $AF(\Theta_{\mathrm{pk}},\Phi_{\mathrm{pk}})$ represents the maximum magnitude of the array factor at the actual peak radiation direction after beam steering. 
$AF^{ref}_{\mathrm{broadside}}$ denotes the reference peak of the magnitude of the array factor at the broadside direction, typically $(\Theta,\Phi)=(0^\circ,0^\circ)$, which is used as the reference level for quantifying scan loss.
The relative gain degradation metric directly quantifies a reduction in the peak magnitude of the array factor resulting from the beam scanning performance relative to the broadside reference response.
Specifically, $G_{\mathrm{rel}}=0$ dB represents the ideal scanning performance with no degradation from the broadside response, whereas $G_{\mathrm{rel}}=-3$ dB corresponds to the 3 dB scanning loss, indicating that the steered beam retains 50\% of the broadside power level. 
Moreover, values approaching $G_{\mathrm{rel}}=-5$ dB indicate more severe degradation but may remain acceptable for applications demanding extremely wide-angle beam coverage. 
Generally, this metric is widely regarded as a practical criterion for acceptable gain degradation resulting from beam scanning in phased-array and metasurface antenna systems. 
For practical beam-steering applications, $G_{\mathrm{rel}}$ values within the range of $-5~\mathrm{dB}\leq G_{\mathrm{rel}}\leq -3~\mathrm{dB}$ are typically regarded as acceptable levels of the gain degradation resulting from beam steering performance.

The SLL quantifies the relative suppression of the undesired radiation with respect to the peak main-lobe response and serves as an important metric for evaluating beamforming quality, interference mitigation, and spatial selectivity. 
\begin{equation}
\mathrm{SLL}
=
20\log_{10}
\left(
\frac{
\displaystyle
\max_{(\Theta,\Phi)\in\Omega_{\mathrm{sidelobe}}}
|AF(\Theta,\Phi)|
}
{|AF(\Theta_{\mathrm{(pk)}},\Phi_{\mathrm{(pk)}})|}
\right)
\label{eq:sll}
\end{equation}
where $\Omega_{\mathrm{sidelobe}}$ represents the angular region excluding the main beam. 
For practical beam-steering applications, SLLs in the range of approximately $-8$ to $-10~\mathrm{dB}$ are considered acceptable for the effective sidelobe suppression.
Consequently, a steering direction is regarded as successfully synthesized when it simultaneously satisfies the four aforementioned criteria.
These metrics quantify not only whether the desired beam direction is achieved but also the quality of the synthesized radiation pattern in terms of pointing accuracy, scan efficiency, and sidelobe suppression.
\subsection{Antenna Efficiency}
The lens transmission efficiency is defined as the average transmission efficiency of the stacked multilayer hemispherical antenna under horn illumination and is expressed as,
\begin{equation}
\eta_{\mathrm{lens}}
\approx
\frac{\displaystyle\sum_{(n,k)}\Delta S_{(n,k)}
|F_{\mathrm{horn},(n,k)}|^{2}
|S_{21,(n,k)}|^{2}}
{\displaystyle\sum_{(n,k)}\Delta S_{(n,k)}
|F_{\mathrm{horn},(n,k)}|^{2}}
\label{eq:eta_lens}
\end{equation}
where $\Delta S_{(n,k)}$ denotes the surface area of the $(n,k)$-th unit cell.
The weighting term accounts for the non-uniform horn illumination across the curved aperture. 
Furthermore, the overall radiation efficiency, which accounts for the total power conversion efficiency of the antenna system, is expressed as,
\begin{equation}
\eta_{\mathrm{rad}}
=
\frac{G}{D}
=
10^{\frac{G_{\mathrm{dBi}}-D_{\mathrm{dBi}}}{10}}
\label{eq:eta_rad}
\end{equation}
where $G$ and $D$ denote the antenna gain and directivity, respectively. 
This metric quantifies the fraction of accepted power effectively radiated by the antenna after considering all dissipative mechanisms, including conductor losses, dielectric losses, and surface-wave losses within the multilayer structure.
It can be extracted directly from CST Studio Suite or computed using simulated gain and directivity.

On the other hand, the mismatch efficiency is defined as,
\begin{equation}
\eta_{\mathrm{mismatch}}
=
1-|S_{11}|^{2}
\label{eq:eta_mismatch}
\end{equation}
which represents the fraction of incident power delivered to the horn input after accounting for power reflected due to impedance mismatch.
Consequently, the total antenna efficiency is expressed as,
\begin{equation}
\eta_{\mathrm{total}}
=
\eta_{\mathrm{mismatch}}
\eta_{\mathrm{rad}}
\label{eq:eta_total}
\end{equation}
It is worth noting that Eq.\ref{eq:eta_total} holds only when $\eta_{\mathrm{rad}}$ excludes the mismatch efficiency contribution.
In CST Studio Suite, the reported realized gain and total efficiency inherently account for the impedance mismatch at the input port. 
Consequently, the mismatch efficiency is already included in these quantities and should not be applied again in subsequent calculations.
Alternatively, when the realized gain is available, the total efficiency can be obtained as,
\begin{equation}
\eta_{\mathrm{total}}
=
\frac{G_{\mathrm{realized}}}{D}
=
10^{\frac{G_{\mathrm{realized,dBi}}-D_{\mathrm{dBi}}}{10}}
\label{eq:eta_total_realized}
\end{equation}
Collectively, these efficiency metrics enable a comprehensive characterization of the power-transfer and radiation performance of our proposed feedhorn hemispherical graphene-based transmitarray antenna, Table~\ref{tab:efficiency_scan}.
\section{CST VBA Macro for Automated Multilayer Hemispherical Transmitarray Modeling}
To facilitate efficient and reproducible electromagnetic modeling of the proposed multilayer hemispherical transmitarray antenna, a dedicated VBA macro was developed in CST Studio Suite.
For the first time in this work, we developed macro to automatically construct the geometry of the stacked multilayer lens, graphene sectors, material assignments, and simulation environment, eliminating repetitive manual modeling and facilitate efficient parametric design and optimization.
\subsection{Macro-based Definition of the Lens Geometry}
Following the validation of the metasurface antenna, without the multilayer stack, illuminated by the circular feedhorn in the CST Studio Suite environment, the custom VBA macro is developed to automate the construction of our proposed multilayer transmitarray antenna.

To automate the model generation process, the macro codes first acquires the geometric, constitutive, and discretization parameters through user-defined InputBox dialogs, Table~\ref{tab:macro_input_parameters}.
Using these inputs, the macro code automatically generates the required CST components, assigns the corresponding material properties, and constructs the complete multilayered transmitarray antenna, including the graphene sectors and other layers in Fig.~\ref{fig:transmitarray_antenna}.
The macros are developed in VBA and executed directly within CST Studio Suite to automate electromagnetic modeling and simulation tasks. 
Our CST VBA macro script automatically constructs the complete 3D electromagnetic model by generating the geometry, assigning the material properties, defining the electromagnetic boundary conditions, and configuring the ports and excitation settings.

Beyond the automated generation of the electromagnetic model, the developed macro codes provide a complete simulation workflow by configuring and controlling the mesh generation process, executing full-wave electromagnetic simulations, and automatically extracting and exporting the required simulation results.

This automated implementation eliminates repetitive manual modeling, ensures consistent and reproducible geometry generation, and significantly reduces model setup time. Moreover, the macro enables efficient parametric studies, optimization, and large-scale design sweeps by automatically regenerating the entire model for different sets of design parameters while maintaining a consistent modeling procedure.

Prior to initiating the geometry generation process, the macro executes an input validation routine to ensure the correctness and physical feasibility of the defined parameters. Specifically, it verifies that the hemispherical radius, dielectric shell thicknesses, and relative permittivity values are positive, while ensuring that the combined thickness of the inner dielectric shells does not exceed the hemispherical radius.
\begin{table}[!t]
\caption{Definition of design parameters used by the CST macro}
\label{tab:macro_input_parameters}
\centering
\small
\renewcommand{\arraystretch}{1.15}
\begin{tabular}{p{3cm} p{5.2cm}}
\hline
\textbf{Design parameter} & \textbf{Corresponding CST model parameter definition} \\
\hline

Hemisphere radius, $R$ &
Radius of the hemispherical lens \\

Cell distribution &
Number of graphene sectors distributed along each ring \\

Inter-cell gap &
Isolation spacing between adjacent graphene sectors \\

Apex cell radius &
Radius of the apex graphene sector \\

Gold-patch radial offset &
Radial offset extending from $R$ to $R+\mathrm{dis}$ and also defines the radial placement offset of the conformal gold patches \\

Inner-shell thickness of dis2 &
Thickness of the inner dielectric shell spanning from $R-\mathrm{dis2}$ to $R$ \\

Additional inner-shell thickness of dis0 &
Radial thickness of the additional inner dielectric shell,
$R-\mathrm{dis2}-\mathrm{dis0}\le r\le R-\mathrm{dis2}$ \\

Additional inner-shell permittivity &
Relative permittivity of the additional inner dielectric shell \\

Inner-shell permittivity &
Relative permittivity of the inner dielectric shell \\

Outer-shell permittivity &
Relative permittivity of the outer dielectric shell \\

\hline
\end{tabular}
\end{table}
\subsection{Dielectric Shells and Layer Interpretation}
The macro automatically generates three concentric, parameterized hemispherical dielectric shell regions and assigns them to the CST components. 
Each shell is constructed through a Boolean subtraction process. 
Two concentric spheres are initially created, and the inner sphere is subtracted from the outer sphere to form a spherical shell. 
A vacuum brick operation is subsequently applied to remove the lower half, resulting in a true hemispherical dielectric shell. 
This approach generates a true hemispherical dielectric shell geometry rather than a complete spherical layer, providing an accurate representation of the conformal lens structure.
It is worth mentioning that the current macro employs generic isotropic dielectric materials for the generated shell regions. 
If the physical multilayer stack is intended to explicitly represent hBN and SiO$_2$, the existing shell-generation methodology can be retained without modification. 
However, these placeholder material definitions should be replaced with the corresponding physical material properties, including the appropriate relative permittivity, loss tangent, and, when required for accurate hBN modeling, anisotropic dielectric tensor parameters.
\subsection{Graphene Sector Generation on the Hemispherical Surface}
The macro script automatically instantiates the graphene sectors within the dedicated component folder in the CST environment.
The macro handles the hemispherical discretization by treating the apex cell separately and then constructing the remaining graphene cells sequentially in a ring-by-ring distribution over the curved surface. This approach ensures accurate conformal placement of the graphene elements while maintaining independent electrical control of each unit cell. This macro-driven layout procedure ensures high-fidelity conformal placement across the hemispherical surface while preserving localized, independent electrostatic bias control for each graphene unit cell.
To introduce an electrical isolation gap between adjacent graphene sectors, the macro slightly shrinks the angular boundaries before constructing each cell according to,
\begin{equation}
\theta^{\mathrm{iso}}_{n} = \theta_n + \frac{\Delta\theta}{2},
\qquad
\theta^{\mathrm{iso}}_{n+1} = \theta_{(n+1)} - \frac{\Delta\theta}{2}
\end{equation}

\begin{equation}
\phi^{\mathrm{iso}}_{(n,k)} = \phi_{(n,k)} + \frac{\Delta\phi}{2},
\qquad
\phi^{\mathrm{iso}}_{(n+1,k+1)} = \phi_{(n+1,k+1)} - \frac{\Delta\phi}{2}
\end{equation}
where $N_n$ and $N_r$ represent the number of rings, excluding the apex unit cell, and the number of unit cells distributed on each ring, respectively.	The azimuthal shrinkage is scaled by the local spherical metric coefficient, 
$\sin(\theta_C)$, where $\theta_C$ represents the central polar angle of the cell. 
This correction ensures a uniform physical isolation gap across the curved hemispherical 
surface by compensating for the variation in azimuthal arc length of 
$L_{\phi}=R\sin(\theta)\Delta\phi$.
In the vicinity of the hemispherical apex, $\sin(\theta)$ approaches zero, potentially causing numerical instability and cell geometry collapse during the discretization process. To avoid this issue, the macro enforces a minimum value for the scaling factor, thereby ensuring robust and accurate generation of apex-adjacent graphene cells.

The main features of the graphene-sector generation procedure are summarized as follows: \textbf{(1) apex graphene cell}: the apex graphene cell is generated independently as a circular polygon conformal to the hemispherical surface. It is discretized using 96 segments to provide a smooth and accurate approximation of the curved boundary while maintaining geometric consistency with the surrounding graphene cells, \textbf{(2) ring graphene sectors}: graphene sectors are generated as four-vertex Polygon3D curves in the spherical coordinate and subsequently converted into planar sheet faces using the CST CoverCurve operation, \textbf{(3) electrical isolation}: gaps are introduced between adjacent graphene sectors to electrically isolate the individual graphene regions, thereby enabling independent electrostatic biasing and programmable control of the local surface impedance, and \textbf{(4) independent material assignment to each graphene sector}: each graphene sector is assigned a dedicated material definition and an independent tuning parameter of $V_{(n,k)}$, thereby enabling independent control of its chemical potential and corresponding electromagnetic response. 
This macro-based generation method enables accurate conformal discretizations of the hemispherical graphene metasurface while preserving the independent tunability required for programmable beam steering and wavefront control.
\subsection{Graphene Material Definition and Chemical Potential Assignment}
For each graphene unit cell, the macro assigns independent tuning parameters of $V_{(n,k)}$  and generates a corresponding graphene material definition. The macro indeed evaluates the Kubo conductivity internally over a wide frequency range and converts the complex surface conductivity into tabulated surface impedance values. 
Graphene material properties are implemented automatically in CST through a macro subroutine that executes four primary tasks: 
\textbf{(1) conductivity evaluation}: it computes graphene surface conductivity over a broad frequency range using the Kubo model, 
\textbf{(2) impedance conversion}: it transforms the complex surface conductivity into equivalent surface impedance values that are tabulated for the CST environment,
\textbf{(3) material creation}: it instantiates the corresponding material definitions within the CST environment, and 
\textbf{(4) property assignment}: it automatically instantiates the CST material definitions and maps the frequency-dependent impedance data directly to the tabulated material properties. 
This formulation provides independent tuning of individual graphene sectors, allowing the hemispherical metasurface to realize spatially inhomogeneous impedance distributions for dynamically programmable wavefront control. 
The main macro parameters associated with the graphene material definition are summarized in Table~\ref{tab:macro_subroutine_variables}. 
\begin{table}[!t]
\caption{Variable quantities within macro subroutines and the macro execution routine}
\label{tab:macro_subroutine_variables}
\centering
\small
\renewcommand{\arraystretch}{1.1}
\begin{tabular}{p{3.1cm} p{5.2cm}}
\hline
\textbf{Variable Quantities within Macro Subroutines} & 
\textbf{Macro Execution Routine} \\
\hline

Voltage-control parameters: 

$V_{(1,1)},\ldots,V_{(N_r, N_n)}$ & 
An independent parameter $V_{(n,k)}$ is assigned to each graphene sector material and directly defines the sector-specific chemical potential used in the Kubo conductivity evaluation \\

Graphene sector material definition & 
Equivalent lossy-metal material definition with tabulated transparent surface impedance values \\

Thickness of each graphene sector & 
Effective graphene thickness parameter used for equivalent permittivity conversion \\

Temperature of each graphene sector & 
The temperature parameter incorporated into the Kubo formulation to evaluate graphene surface conductivity \\

Carrier relaxation time & 
Carrier relaxation-time parameter $\tau$ used to model scattering losses in the graphene Kubo conductivity formulation \\

Chemical potential of each graphene sector & 
The chemical potential parameter is directly scaled by $V_{(n,k)}$ \\
\hline
\end{tabular}
\end{table}
Although the macro labels voltage variables as $V_{(1,1)},\ldots,V_{(N_r,N_n)}$, our implementation directly interprets these parameters as the graphene chemical potentials expressed in electron volts. 
Consequently, $V_{(n,k)}$ represents the chemical potential of the $i$-th graphene sector rather than an applied voltage $V_{(n,k)} \equiv \mu_{c,(n,k)}$. A voltage-controlled interpretation would require an additional electrostatic model to establish the relationship between the gate bias $V_g$ and the induced chemical potential
$\mu_c(V_g)$. If the final physical model is intended to describe voltage-controlled graphene tuning, the macro should either rename these parameters as $\mu_{c,(n,k)}$ or incorporate the electrostatic relationship between gate voltage and chemical potential using the gate capacitance of the dielectric layer (e.g., hBN-based gating structure). 
The intra-band term dominates at our operating frequency of 250 GHz and typical chemical potentials of 0.2–0.8 eV, the implemented full Kubo-type numerical evaluation ensures accuracy and validity across a broader terahertz (THz) spectrum.
\subsection{Gold Patch Placement on the Hemispherical Surface}
The macro employs a predefined reference metallic patch as a template for generating the corresponding elements across the aperture. 
For each unit cell, the reference geometry is replicated and subsequently scaled, rotated, and translated according to the corresponding cell dimensions and spatial coordinates, thereby constructing the successive active rows. 

\textbf{(1) Prerequisite}:
prior to executing the macro, a reference gold patch should be defined within the active CST project. This reference patch serves as the template for generating the conformal gold-patch array and should exist in the model before the macro is executed.

\textbf{(2) Algorithmic flow}: the macro generates the gold-patch distribution by creating one copy of the reference patch for each graphene sector. Each copied patch subsequently undergoes a sequence of geometric transformations to achieve conformal placement on the hemispherical surface: \textbf{duplication}: a copy of the reference patch is created, \textbf{scaling}: the copied patch is resized according to the prescribed scaling factor, \textbf{orientation}: the patch is rotated about the local coordinate axes to align with the target graphene sector, where the rotation angles are determined by the corresponding spherical coordinates of $(\theta,\phi)$, and \textbf{positioning}: the transformed patch is translated to the cartesian coordinate system corresponding to the center of the target graphene sector.

\textbf{(3) Spatial mapping}: the spatial placement of the gold patches is determined according to the hemispherical cell distribution: \textbf{apex cell}: the gold patch associated with the apex graphene cell is positioned at the hemispherical apex using the default spherical coordinates $(\theta,\phi)=(0,0)$, \textbf{ring cells}: for all non-apex graphene sectors, each gold patch is positioned at the center of the corresponding graphene sector using its spherical coordinates of $(\theta_C,\phi_C)$, and \textbf{radial placement}: the patch center is evaluated at a radial distance $r=R+\mathrm{dis}$, ensuring that the gold patch is located above the graphene layer and dielectric spacer while preserving the intended multilayer configuration, and \textbf{scaling law and parametric control}: the default implementation employs geometric scaling across successive rings to the patch dimensions,
$\mathrm{patchScale}=0.8^{\mathrm{ringIndex}}$, resulting in a gradual reduction in patch size with increasing ring index to account for the spatially varying local geometry of the hemispherical surface.

The graphene conductivity is initially calculated, and the corresponding tabulated surface-impedance and equivalent-permittivity materials are created during the generation of the graphene sectors. After the resonator patches are added to the previous configuration, the existing tabulated surface-impedance and equivalent-permittivity materials are updated to account for the modified unit cells.
\subsection{Macro Architecture and Subroutine Functionality}
The CST VBA macro features a modular architecture comprised of dedicated subroutines, each executing a distinct phase of the hemispherical metasurface construction. This design enhances code readability, simplifies maintenance, facilitates future feature extensions, and allows independent validation of geometry-generation and material-definition routines. The core subroutines and their specific operational roles are summarized in Table~\ref{tab:macro_subroutines}. 
\begin{table}[!t]
\caption{Generalized functions of the CST VBA macro subroutines}
\label{tab:macro_subroutines}
\centering
\small
\renewcommand{\arraystretch}{1.1}
\begin{tabular}{p{3.0cm} p{5.3cm}}
\hline
\textbf{Subroutine} & \textbf{Functionality} \\
\hline
Main &
Collects user inputs, validates design parameters, generates the antenna geometry, assigns material properties, and constructs the conformal multilayer model. \\

Input parser &
Converts user-defined cell distributions into numerical arrays specifying the number of unit cells in each hemispherical ring. \\

Material definition &
Creates dielectric and graphene material models using the specified constitutive parameters. \\

Dielectric shell generation &
Constructs concentric hemispherical dielectric shells from user-defined geometric parameters. \\

Coordinate transformation &
Transforms spherical coordinates into Cartesian coordinates for geometry generation and object placement. \\

Apex cell generation &
Constructs the graphene unit cell located at the hemispherical apex. \\

Graphene cell generation &
Generates conformal graphene unit cells on the hemispherical surface according to the prescribed discretization. \\

Gold patch placement &
Replicates, scales, rotates, and positions the reference gold patch at the center of each graphene sector. \\

Graphene material modeling &
Evaluates the graphene conductivity using the Kubo model and generates the corresponding frequency-dependent equivalent material properties. \\
\hline
\end{tabular}
\end{table}
\subsection{Horn and Hemisphere Alignment}
After generating the hemispherical lens geometry, the horn antenna is positioned such that its phase center is located at, or very close to, the origin of the hemispherical coordinate system. 
This center-fed configuration ensures uniform illumination because the distance from the horn's phase center to the hemispherical aperture remains approximately constant. Consequently, the incident spherical wavefront maintains a nearly uniform phase distribution across the lens surface, minimizing the phase compensation required by the metasurface elements.
To ensure uniform and symmetric illumination of the aperture, the horn boresight should be aligned with the +z-axis when the hemispherical lens aperture is oriented toward the positive z-direction.
In the CST assembly model, the hemispherical lens should be positioned within the far-field region of the horn antenna. 
The far-field criterion establishes the initial horn-to-lens separation, while the near-field phase map verifies that the electromagnetic field illuminating the hemispherical lens maintains sufficient phase and amplitude uniformity. 
Indeed, under the ideal assumption of a uniform incident phase, 
the required transmission phase for the $(n,k)$-th unit cell is given by,
\begin{equation}
\psi_{\mathrm{req},(n,k)}
=
-k_0\,
\hat{\mathbf{s}}_0
\cdot
\mathbf{r}_{(n,k)}
+C
\label{eq:required_phase}
\end{equation}
where $k_0$, $\hat{\mathbf{s}}_0$, and $\mathbf{r}_{(n,k)}$ denote the free-space wavenumber, the desired beam-steering direction, and the position vector of the $(n,k)$-th unit cell, respectively. The constant of $C$ represents an arbitrary reference phase.
In practice, however, the incident field radiated by the horn antenna possesses a spatial variation in its phase distribution,
\begin{equation}
\psi_{\mathrm{incident},(n,k)}
=
\angle
E_{\mathrm{inc}}
(\mathbf{r}_{(n,k)})
\label{eq:incident_phase}
\end{equation}
where $E_{\mathrm{inc}}(\mathbf{r}_{(n,k)})$ denotes the complex incident electric field at the location of the $(n,k)$-th unit cell. Consequently, the total transmitted phase includes the phase introduced by the corresponding graphene unit cell,
$\psi_{\mathrm{trans},(n,k)}
=
\angle S_{21,(n,k)}$,
yielding,
\begin{equation}
\psi_{\mathrm{total},(n,k)}
=
\psi_{\mathrm{incident},(n,k)}
+
\psi_{\mathrm{trans},(n,k)}
\label{eq:total_phase}
\end{equation}
Therefore, the transmitted wavefront no longer matches the desired phase distribution, causing beam-pointing errors, degraded beam-steering performance, elevated sidelobe levels, and a reduction in the achievable antenna gain.
To overcome this limitation, and thereby achieve constructive interference along the desired beam-steering direction, the transmitted phase profile should incorporate the actual incident wavefront distribution, i.e., Eq.\ref{eq:incident_phase}, and satisfy the near-field phase-map condition,
\begin{equation}
\psi_{\mathrm{incident},(n,k)}
+
\psi_{\mathrm{trans},(n,k)}
\approx
\psi_{\mathrm{req},(n,k)}
\label{eq:phase_matching}
\end{equation}
This phase distribution defines the required transmission phase response of the programmable graphene-based unit cells to achieve the desired beam-steering functionality.
For CST verification of the near-field phase map over the hemispherical coordinates, first the minimum horn-to-lens distance is estimated using the far-field (Fraunhofer) criterion,
\begin{equation}
d
\ge
\frac{2D_{\mathrm{horn}}^{2}}{\lambda}
\label{eq:fraunhofer}
\end{equation}
where $D_{\mathrm{horn}}$ denotes the maximum aperture dimension of the feed horn antenna. The hemispherical lens is then positioned at the calculated minimum horn-to-lens distance within the CST Studio Suite environment.
Subsequently, the phase and amplitude of the incident electric field are evaluated at the physical lens position as follows,
\begin{equation}
E(\theta,\phi)
=
|E(\theta,\phi)|
e^{j\psi_{inc}(\theta,\phi)}
\label{eq:incident_field}
\end{equation}
where $|E(\theta,\phi)|$ denotes the incident field amplitude and
$\psi_{\mathrm{incident}}
=
\psi_{inc}(\theta,\phi)
=
\angle E(\theta,\phi)$
represents the corresponding incident field phase distribution over the hemispherical aperture.
The near-field verification in the CST Studio Suite environment can be fulfilled by the two criteria of phase curvature and amplitude uniformity.
The amplitude uniformity is evaluated by the spatial variation of the field amplitude across the aperture given by,
\begin{equation}
\Delta A
=
\frac{
\max |E(\theta,\phi)|
-
\min |E(\theta,\phi)|
}{
\max |E(\theta,\phi)|
}
\label{eq:amplitude_uniformity}
\end{equation}
The phase curvature measures the total phase variation introduced by the horn wavefront across the lens aperture,
\begin{equation}
\Delta\psi_{\mathrm{incident}}
=
\max\!\left(\psi_{inc}(\theta,\phi)\right)
-
\min\!\left(\psi_{inc}(\theta,\phi)\right)
\label{eq:phase_curvature}
\end{equation}
The horn illumination should be verified to ensure consistency with the assumptions of the phase-synthesis formulation. If significant phase curvature or nonuniform amplitude illumination persists across the hemispherical aperture, the horn-to-lens distance should be increased or adjusted until the required illumination conditions are satisfied.
\subsection{Hybrid Solver Strategy in the CST Environment}
For our transmitarray antenna, the wavelength propagating inside the SiO$_2$ dielectric layer is given by,
\begin{equation}
\lambda_{\mathrm{SiO_2}}
=
\frac{\lambda_0}{\sqrt{\varepsilon_r}}
\label{eq:lambda_sio2}
\end{equation}
$\varepsilon_r$ represents the relative permittivity of the SiO$_2$ layer. Hence, the effective wavelength inside SiO$_2$ becomes,
\begin{equation}
\lambda_{\mathrm{SiO_2}}
=
\frac{\lambda_0}{\sqrt{3.9}}
\approx
0.506\lambda_0
\label{eq:lambda_sio2_value}
\end{equation}
For both frequency- and time-domain full-wave solvers in CST Studio Suite, the maximum mesh size should satisfy electromagnetic discretization requirements, and is typically selected to be smaller than $\lambda/5$, finer resolutions like $\lambda/10$ or $\lambda/20$ are used for higher accuracy. Furthermore, in CST Studio Suite frequency-domain simulations, the tetrahedral and hexahedral mesh resolutions are typically set to $\lambda/4$, and $\lambda/8$, respectively.
The maximum allowed tetrahedral mesh edge length under a refined $\lambda/10$ resolution criterion is chosen by the following equation,
\begin{equation}
\Delta_{\mathrm{mesh,tet}}
\le
\frac{\lambda_{\mathrm{SiO_2}}}{4\times10}
=
\frac{0.506\lambda_0}{40}
\approx
0.0126\lambda_0
\label{eq:tet_mesh}
\end{equation}
where $\Delta_{\mathrm{mesh,tet}}$ denotes the maximum allowed tetrahedral mesh edge length. Therefore, for our transmitarray antenna operating at 250 GHz, the maximum tetrahedral mesh edge length based on this refined criterion is approximately given by,
\begin{equation}
\Delta_{\mathrm{mesh,tet}}
\le
15.1~\mu\mathrm{m}
\label{eq:tet_mesh_250GHz}
\end{equation}
Moreover, nanometer-scale graphene layers, hBN dielectric layers, and a large number of conformal gold resonators can become computationally prohibitive due to the extreme multiscale nature of the problem. Consequently, our simulation constitutes an electrically large electromagnetic problem, particularly within the region between the horn antenna and the lens; conversely, the graphene, hBN, and gold resonator patches are much smaller and locally resonant. Its computational complexity is comparable to the full-wave analysis of large-scale platforms, such as maritime vessels with extensive radar cross-sections (RCS).

One way to overcome this constraint and substantially reduce computation time is to implement a hybrid solver strategy with a unidirectional assembly of the horn antenna and lens. Our simulation procedure begins by independently modeling and simulating the horn antenna using the time-domain solver. Subsequently, the hemispherical lens, consisting solely of the SiO$_2$ substrate, is modeled and analyzed using the frequency-domain solver.

CST Studio solves the coupled model of the horn antenna and the hemispherical lens including only the SiO$_2$ substrate. The computed electromagnetic field distribution is stored and subsequently used as a new excitation source for the subsequent simulation stage. Rather than approximating the excitation as an ideal plane wave, the saved field solution preserves the realistic illumination characteristics of the horn antenna, including its amplitude taper and phase curvature across the hemispherical aperture.

The assembly of these two components can be performed using either a unidirectional coupling approach or a bidirectional coupling approach, depending on the desired level of electromagnetic interaction and computational accuracy. Since the lens is positioned within the far-field region of the horn antenna, multiple-scattering and backscattering effects are considered negligible. Therefore, a unidirectional coupling approach is adopted for our simulation. The remaining transmitarray layers are then added sequentially, and the simulation process is repeated after the incorporation of each layer until the complete multilayer transmitarray antenna is obtained by advancing the simulation in a forward-only manner, without returning to or re-solving the preceding stages.

Generally, for an initial engineering workflow, a unidirectional coupling approach provides a practical, computationally efficient first step. This method is particularly suitable when the lens introduces negligible perturbation or detuning to the horn antenna. However, if significant back-reflection, impedance mismatch, or strong near-field interactions are observed, a bidirectional coupling scheme or a full-wave analysis of the entire system should be employed to ensure modeling accuracy, Table~\ref{tab:hybrid_simulation_workflow}.
\begin{table*}[!t]
\caption{Hybrid solver of the CST environment for the proposed multilayer transmitarray antenna}
\label{tab:hybrid_simulation_workflow}
\centering
\small
\renewcommand{\arraystretch}{1.15}
\begin{tabular}{p{0.51cm} p{4.6cm} p{4.9cm} p{7cm}}
\hline
\textbf{Stage} & \textbf{CST Task} & \textbf{Solver} & \textbf{Purpose} \\
\hline

1 &
Horn antenna and lens including only a smooth hemispherical SiO$_2$ shell &
A transient solver is used for the horn antenna, while a frequency-domain or asymptotic platform solver models the smooth dielectric shell &
Compute the horn radiation pattern and determine the incident electromagnetic field distribution illuminating the hemispherical lens aperture. \\

2 &
Assembly coupling &
Configure unidirectional hybrid solver for the horn--lens assembly &
Couple the horn source task with the smooth lens platform task without requiring the simultaneous meshing of all fine transmitarray features. \\

3 &
Electromagnetic field export &
Store the computed near-field distribution or field-monitor data at the operating frequency of 250~GHz for subsequent use as the excitation source &
Generate a reusable field excitation that serves as a new source for the subsequent layers in the transmitarray antenna. \\

4 &
Complete multilayer transmitarray architecture incorporating graphene sectors, gold FCC unit cells, and hBN interfacial layers &
A frequency- or time-domain solver driven by field excitations imported from prior simulation stages &
Evaluate the impact of the voltage-controlled graphene sectors and gold patches on the transmission coefficient $S_{21}$, transmission phase response, and far-field beam-steering performance of the transmitarray antenna. \\

\hline
\end{tabular}
\end{table*}
\subsection{Electromagnetic Field Preservation and Reuse}
In the proposed hybrid solver, the horn antenna is first simulated independently; the resulting electromagnetic field distribution at the lens interface is saved and subsequently imported as an equivalent excitation source.
The computed field distribution from the horn antenna and lens including only SiO$_{2}$, as a new excitation source, can then be imported into a second simulation containing the detailed graphene sectors and gold patches. This approach eliminates the need to repeatedly solve the horn antenna for each graphene voltage sector, significantly reducing the computational cost of voltage-sweep simulations.
The complex electric and magnetic near-field distributions should be exported while maintaining a consistent phase reference for subsequent field excitation and coupling analysis.

The same coordinate system should be used, reference origin, and lens position when importing the field distribution into the detailed unit-cell model to ensure accurate phase and amplitude alignment.
We should verify that the polarization of the imported field is consistent with the local orientation of the metallic patches and the principal direction of the graphene conductivity tensor.
Once imported, the electromagnetic field distribution serves as the excitation source for detailed transient or frequency-domain simulation of the graphene–gold patterned lens.
Evaluate the resulting far-field radiation patterns across multiple bias voltage matrices to demonstrate the dynamic beam-steering capability of our tunable transmitarray antenna.
Although this approach significantly reduces computational cost, the accuracy of the imported field should be verified; any discrepancies in the extracted amplitude or phase distribution directly degrade the synthesized transmitarray aperture field and beam-steering performance.
This verification step ensures that the excitation from the imported field accurately reproduces the physical illumination of the original horn antenna prior to executing the computationally expensive graphene voltage-sweep simulations.

Following the successful validation, the imported field executes across the entire graphene-based transmitarray model for the required gate-voltage and chemical-potential sweeps.
This ensures that the observed beam-steering behavior is caused by the graphene tunability rather than numerical artifacts from the imported field process.
\begin{figure}[!t]
\centering
\begin{minipage}[t]{0.48\columnwidth}
\centering
\includegraphics[width=\linewidth]{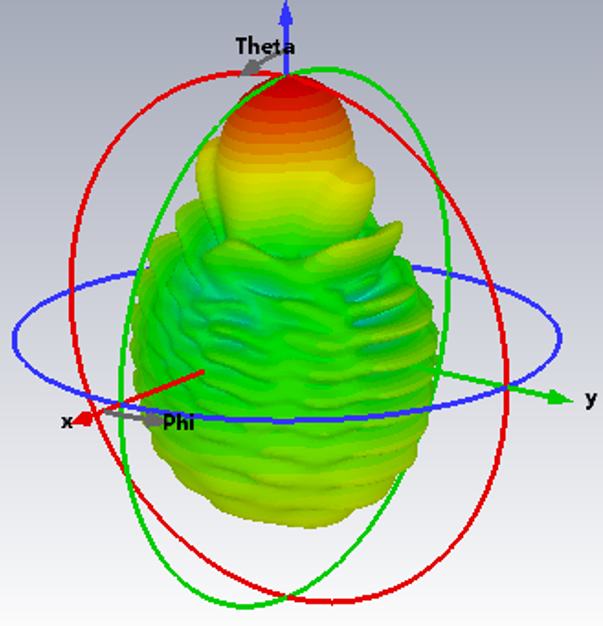}
(a)
\end{minipage}
\hfill
\begin{minipage}[t]{0.48\columnwidth}
\centering
\includegraphics[width=\linewidth]{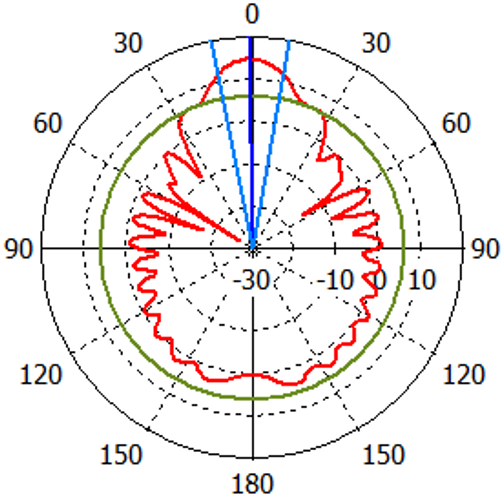}
(b)
\end{minipage}
\vspace{1.5mm}
\begin{minipage}[t]{0.48\columnwidth}
\centering
\includegraphics[width=\linewidth]{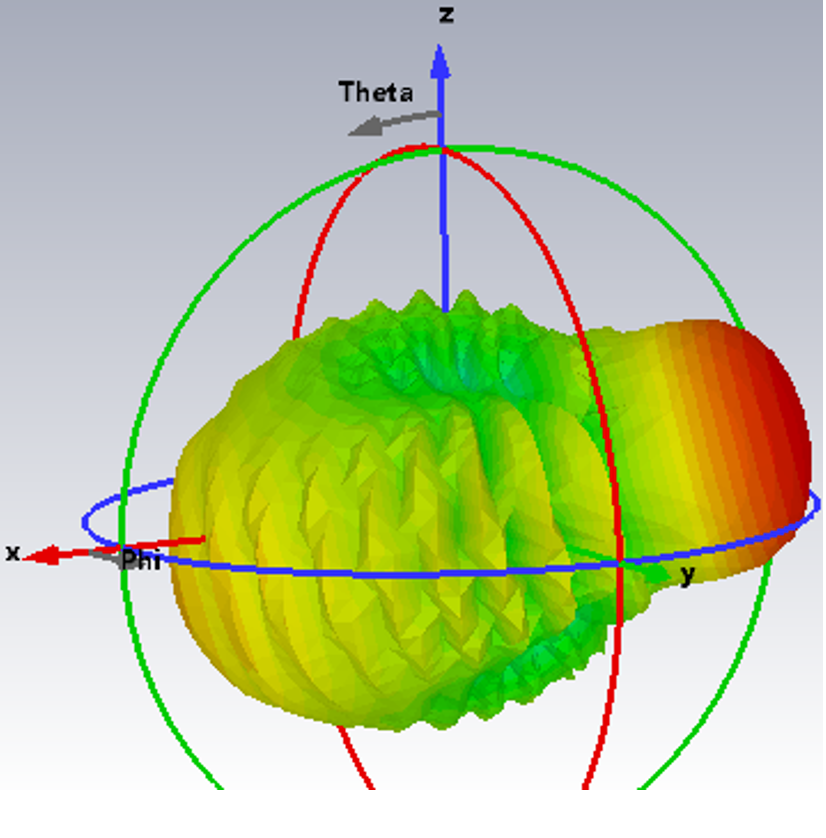}
(c)
\end{minipage}
\hfill
\begin{minipage}[t]{0.48\columnwidth}
\centering
\includegraphics[width=\linewidth]{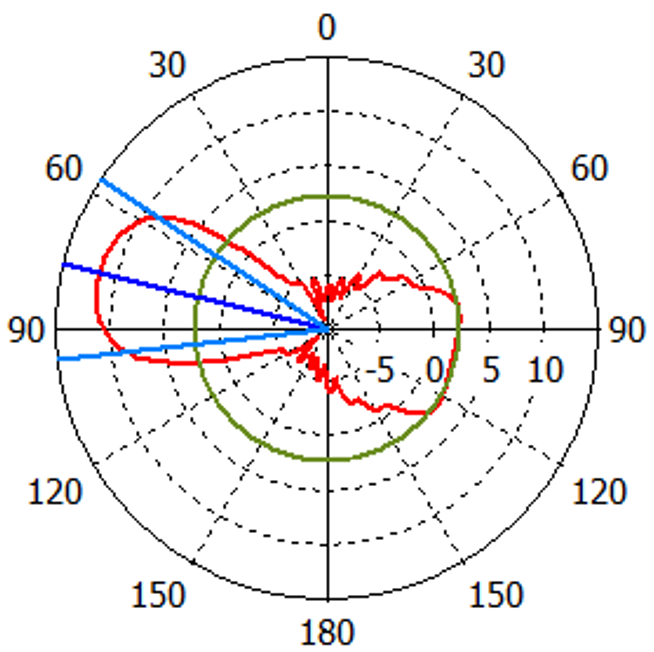}
(d)
\end{minipage}
\vspace{1.5mm}
\begin{minipage}[t]{0.48\columnwidth}
\centering
\includegraphics[width=\linewidth]{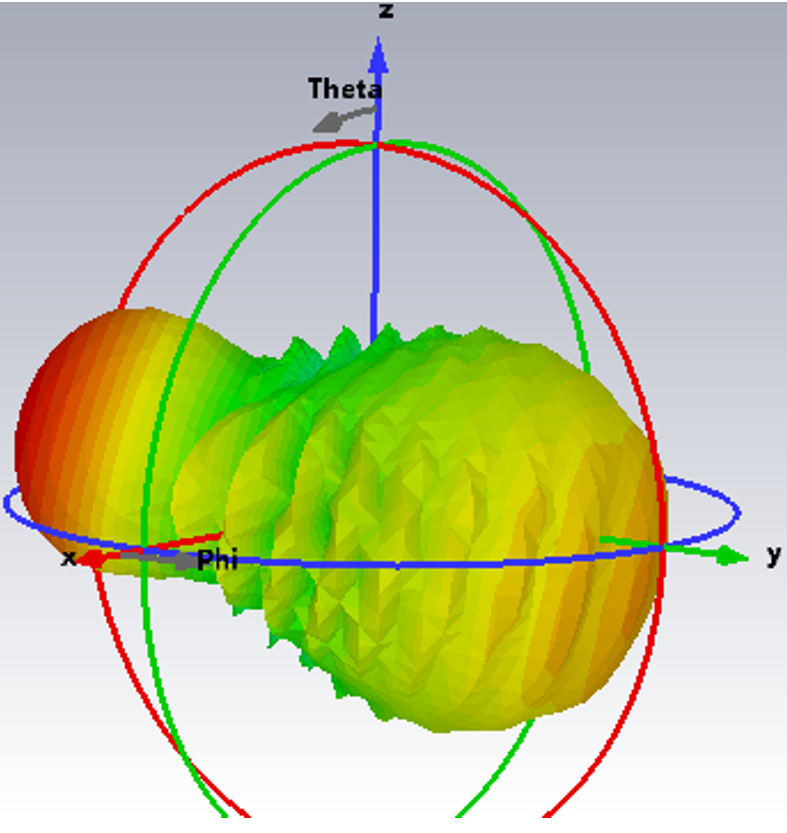}
(e)
\end{minipage}
\hfill
\begin{minipage}[t]{0.48\columnwidth}
\centering
\includegraphics[width=\linewidth]{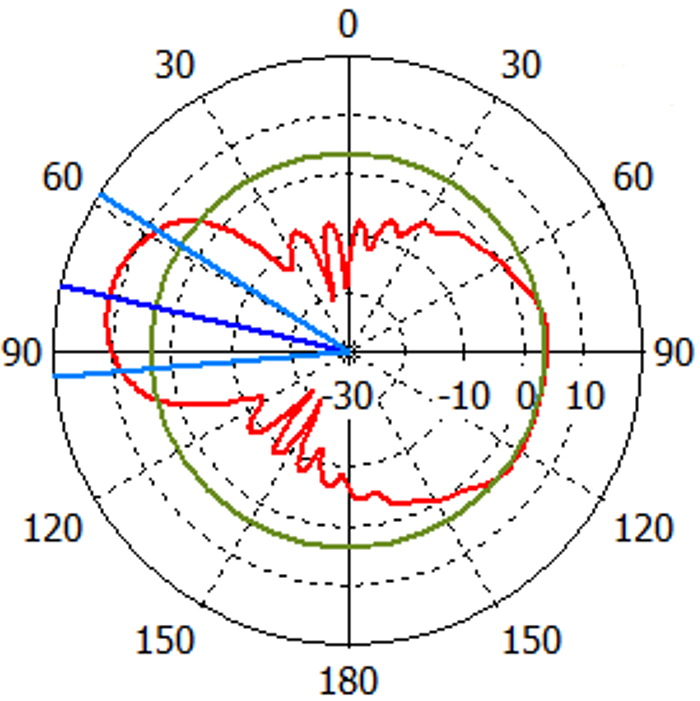}
(f)
\end{minipage}
\vspace{1.5mm}
\begin{minipage}[t]{0.48\columnwidth}
\centering
\includegraphics[width=\linewidth]{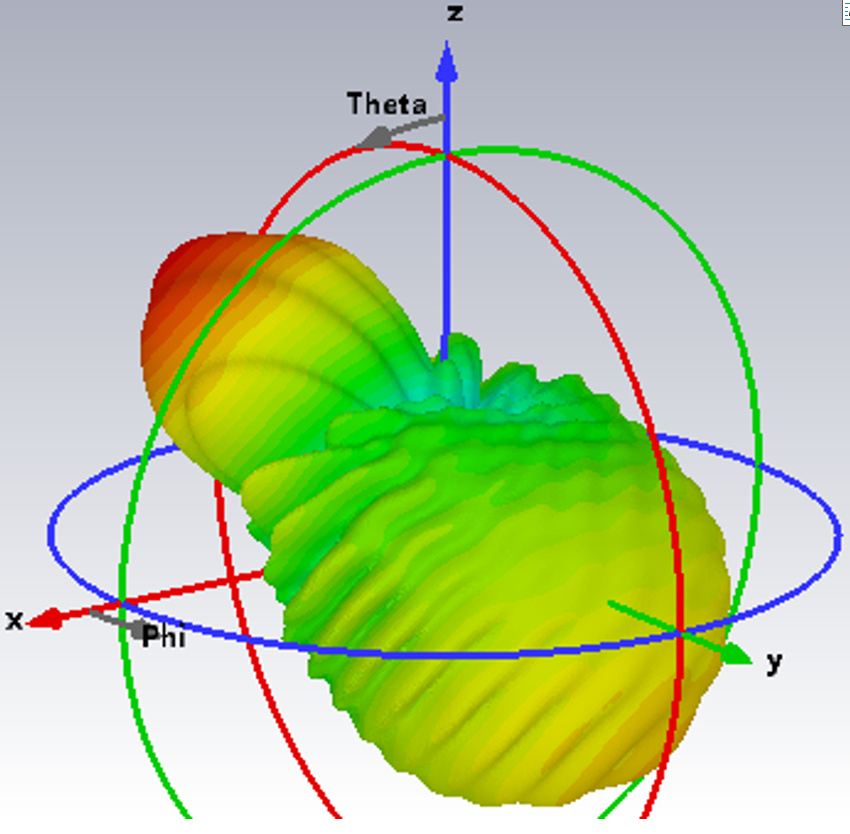}
(g)
\end{minipage}
\hfill
\begin{minipage}[t]{0.48\columnwidth}
\centering
\includegraphics[width=\linewidth]{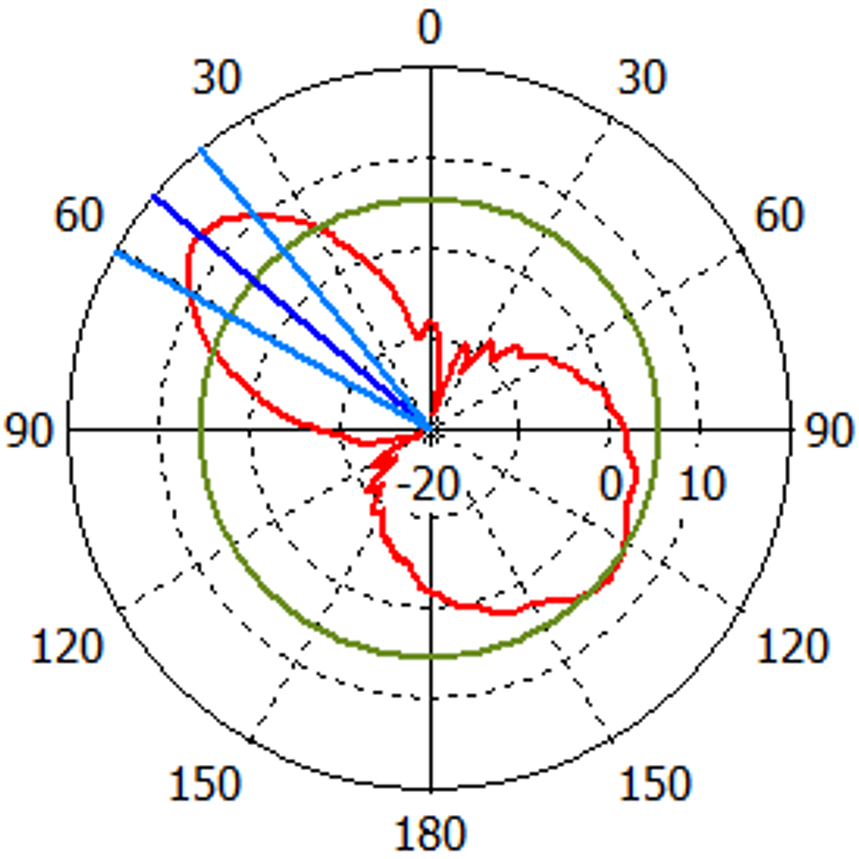}
(h)
\end{minipage}
\end{figure}
\begin{figure}[!t]
\caption{
Demonstration of the wide-angle beam-steering performance of the proposed transmitarray antenna, Fig.~\ref{fig:transmitarray_antenna}, through 3D and 2D radiation patterns. 
The 2D radiation patterns are plotted in the $\phi=90^\circ$ plane as a function of the elevation angle of $\theta$ and the main-lobe magnitude in dBi. 
The proposed transmitarray antenna provides a full azimuthal scanning range of $\phi=360^\circ$ and an elevation scanning range from $\theta=-78^\circ$ to $+78^\circ$ at an operating frequency of 250~GHz. 
The beam-steering directions are illustrated for representative steering angles: 
(1) for $\theta=0^\circ$ and $\phi=0^\circ$: (a) 3D radiation pattern with a directivity of 14.72~dBi and (b) 2D radiation pattern with a HPBW of $21.6^\circ$, 
(2) for $\theta=76^\circ$ and $\phi=150^\circ$: (a) 3D radiation pattern with a directivity of 11.70~dBi and (b) 2D radiation pattern with with a HPBW of $40.1^\circ$,  
(3) for $\theta=78^\circ$ and $\phi=360^\circ$: (a) 3D radiation pattern with a directivity of 11.38~dBi and (b) 2D radiation pattern with a HPBW of $41.8^\circ$, and 
(4) for $\theta=50^\circ$ and $\phi=150^\circ$: (a) 3D radiation pattern with a directivity of 13.24~dBi and (b) 2D radiation pattern with a HPBW of $21.3^\circ$.}
\label{fig:scanning_angles}
\end{figure}

\begin{figure}[H]
\centering
\subfloat[]{%
\includegraphics[width=0.98\columnwidth,height=3.68cm]{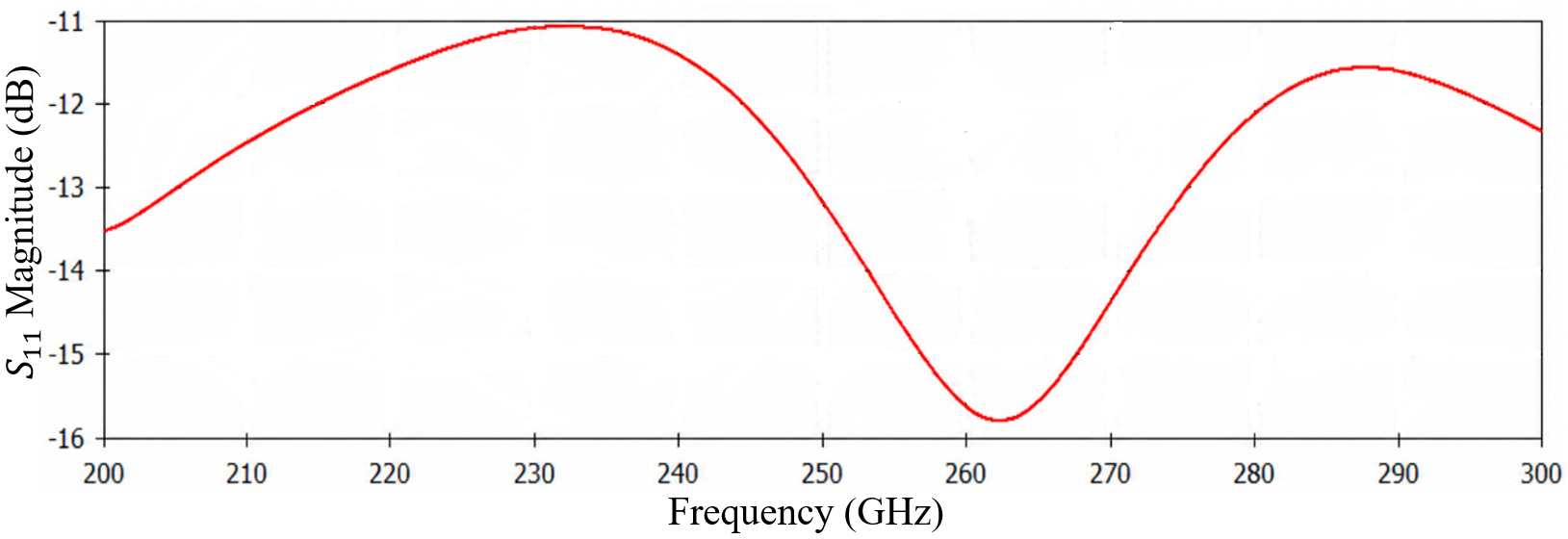}
\label{fig:sub1}
}\\[2mm]

\subfloat[]{%
\includegraphics[width=0.98\columnwidth,height=3.68cm]{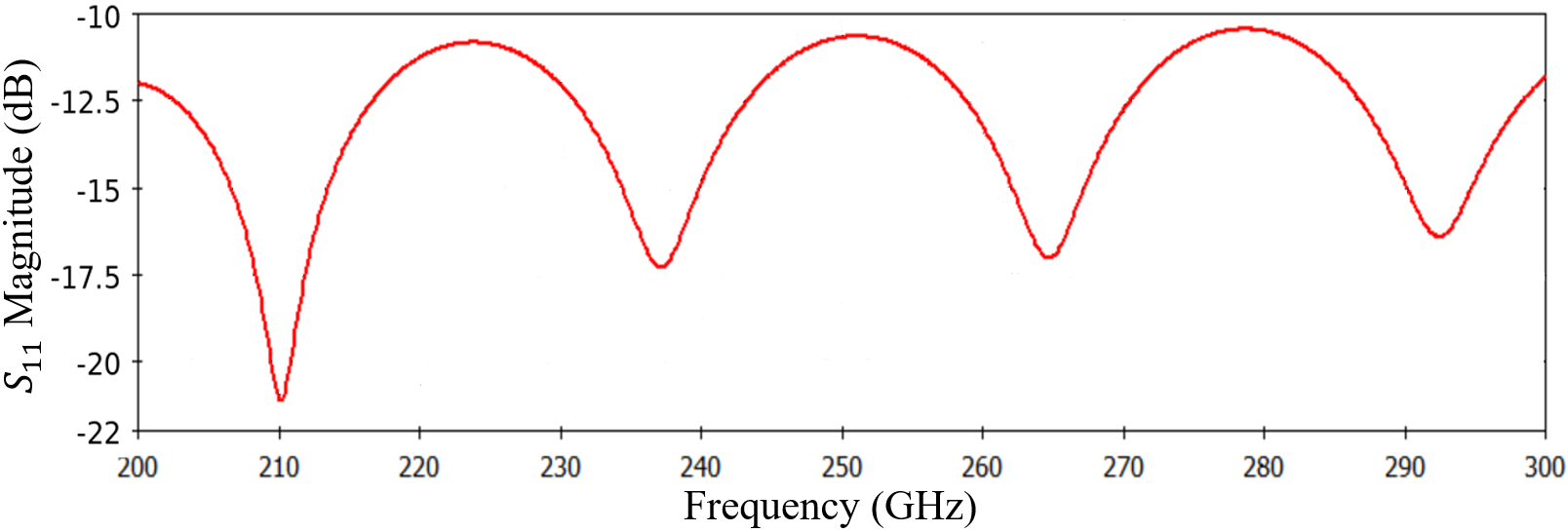}
\label{fig:sub2}
}\\[2mm]

\subfloat[]{%
\includegraphics[width=0.98\columnwidth,height=3.68cm]{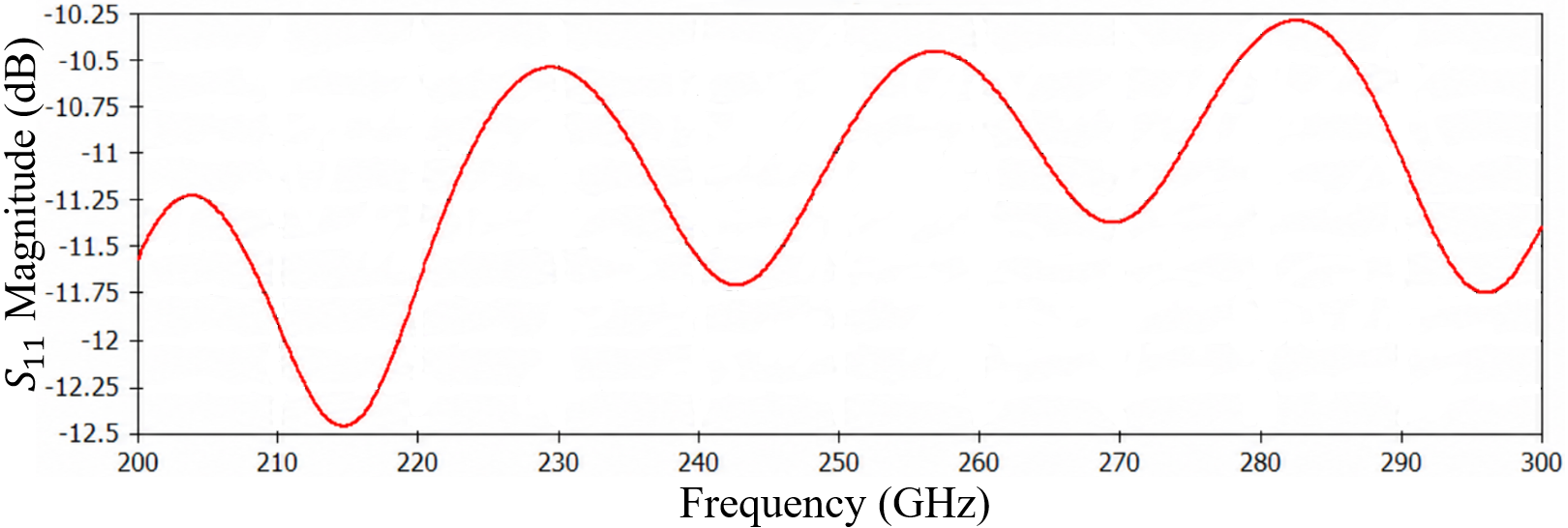}
\label{fig:sub3}
}\\[2mm]
\end{figure}

\begin{figure}[H]
    \centering
    \setcounter{subfigure}{3}

    \subfloat[]{%
        \includegraphics[width=0.98\columnwidth,height=3.68cm]{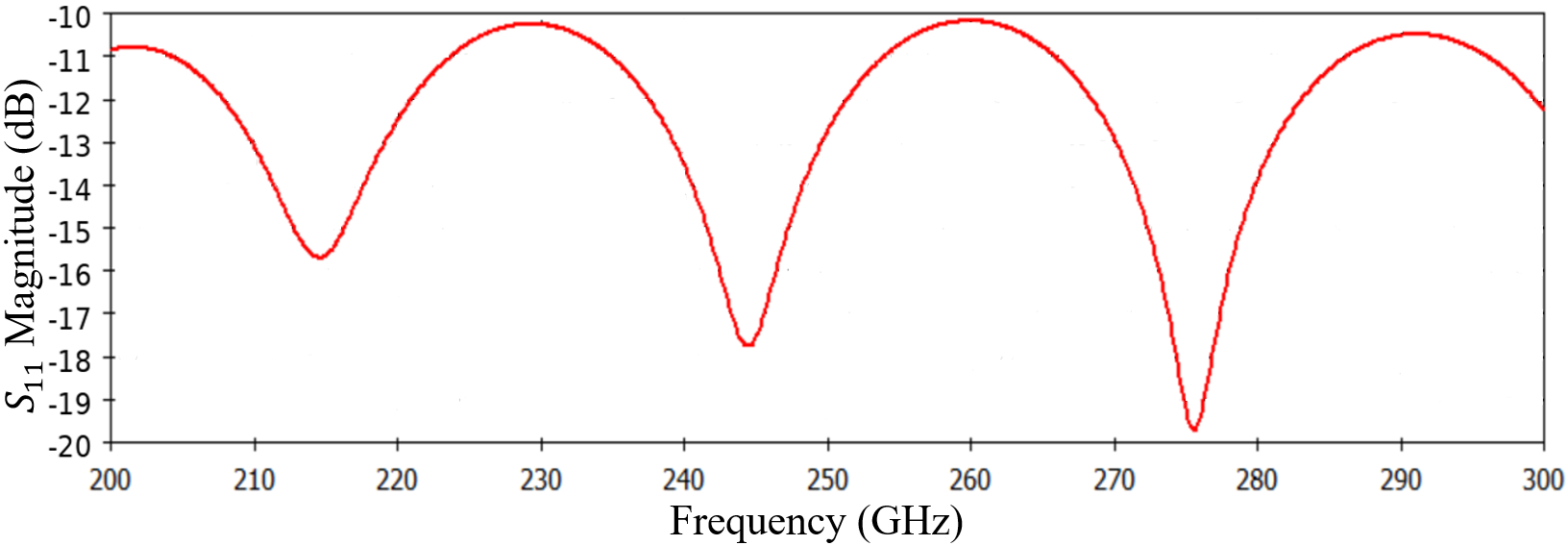}
        \label{fig:sub4}
    }\\[2mm]
    \subfloat[]{%
        \includegraphics[width=0.98\columnwidth,height=3.68cm]{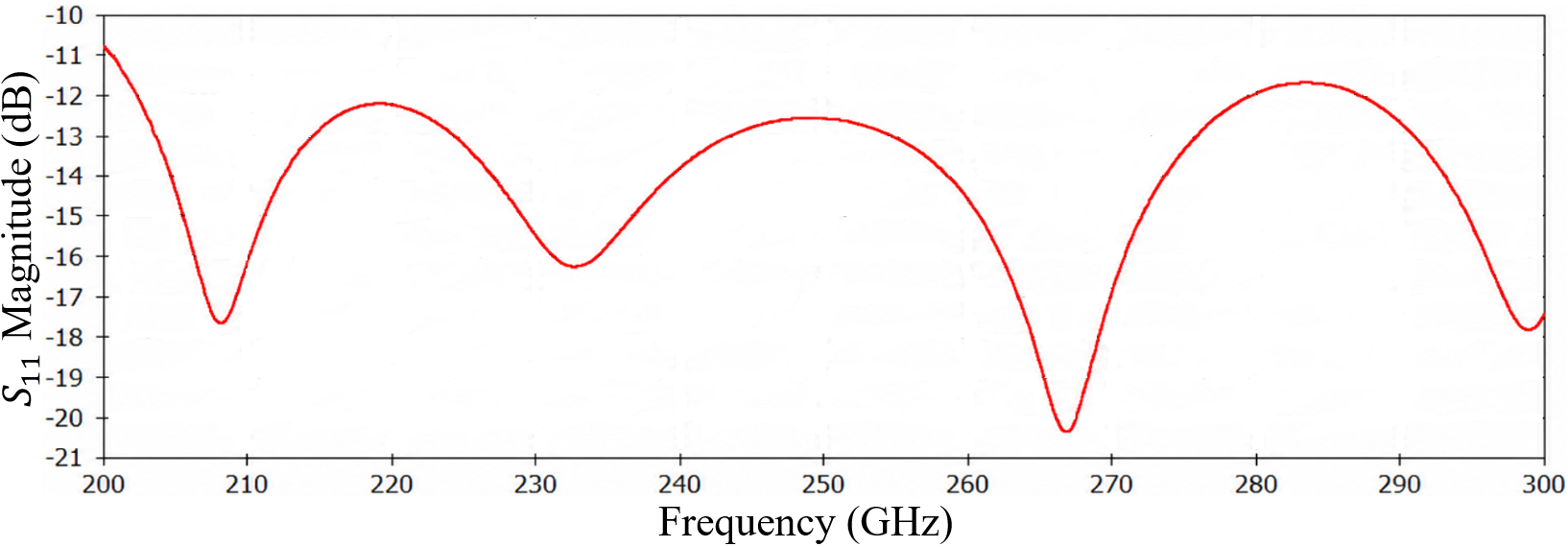}
        \label{fig:sub5}
    }
    \caption{Variation of $S_{11}$ magnitudes in terms of frequency for our proposed transmitarray antenna, consistent with Fig.~\ref{fig:transmitarray_antenna}, for different beam-steering directions of (a) $\theta=0^\circ$ and $\phi=0^\circ$, (b) $\theta=23^\circ$ and $\phi=300^\circ$, (c) $\theta=50^\circ$ and $\phi=150^\circ$, (d) $\theta=78^\circ$ and $\phi=150^\circ$, and (e) $\theta=77^\circ$ and $\phi=300^\circ$}
    \label{fig:five_rows}
\end{figure}
To validate the hybrid solver, both configurations should yield nearly identical fields,
\begin{equation}
E_{\mathrm{direct}}(\theta,\phi)
\approx
E_{\mathrm{imported}}(\theta,\phi)
\label{eq:field_comparison}
\end{equation}
\begin{equation}
AF_{\mathrm{direct}}(\theta,\phi)
\approx
AF_{\mathrm{imported}}(\theta,\phi)
\label{eq:af_comparison}
\end{equation}
The validation procedure compares two equivalent simulation setups: (1) a reference simulation involving direct horn excitation, where the smooth hemispherical SiO$_2$ lens is modeled alongside the original horn antenna. The horn directly illuminates the SiO$_2$-only lens, and the resulting transmitted near-field distribution and far-field radiation patterns are recorded, and (2) an imported-field simulation, where the horn antenna is removed, and the same smooth lens is excited using the previously saved electromagnetic field distribution obtained from the horn simulation. The corresponding transmitted near-field and far-field results are then extracted.

Good agreement confirms that the imported field preserves the essential characteristics of the horn illumination, including amplitude taper across the hemispherical aperture, incident phase curvature, polarization distribution, and spatial phase reference. 
\begin{itemize}
\item Export Source Field: Save the 250 GHz electromagnetic field distribution generated by the horn antenna illuminating the hemispherical lens containing only the SiO$_2$ substrate.
\item Import Nearfield: Import the extracted electromagnetic field distribution as a new excitation source for the remaining graphene--gold transmitarray antenna.
\item Run Full Sim: Execute the final simulation using the imported source excitation.
\item Extract Metrics: Record the reflection coefficient $S_{11}$ and transmission coefficient $S_{21}$ of the transmitarray antenna.
\item Analyze Radiation: Run the voltage-matrix configurations and extract the corresponding far-field radiation patterns, beam peak direction, realized gain, SLL, and horn reflection coefficient $S_{11}$.
\end{itemize}

\subsection{CST Simulation Workflow}
\textbf{Phase 1: Base Geometry and Macro Execution}
\begin{itemize}
\item Initialize Patch: Place Patch17 in Component2 at the coordinate origin.
\item Verify Properties: Confirm the correct material assignment, thickness, and spatial orientation.
\item Run Macro: Execute the script to generate the parametric transmitarray layout.
\item Input Parameters: Specify the hemisphere radius, the number of cells per ring, gap thickness, and top-cell radius.
\item Define Layer Specs: Enter the exact dielectric thicknesses and material permittivities.
\end{itemize}
\textbf{Phase 2: Material and Bias Assignment}
\begin{itemize}
\item Audit Structures: Verify the generated Dielectric\_Shells, Graphene\_Cells, and Patch17 shapes.
\item Assign Dielectrics: Apply specific properties for hBN and SiO$_2$ layers.
\item Configure Graphene: Set up the chemical potentials of $\mu_c$ for graphene sectors.
\item Map Voltages: Implement the graphene bias parameters of $V_{(1,1)},\ldots,V_{(N_r, N_n)}$ using the corresponding voltage-to-$\mu_c$ conversion model.
\end{itemize}
\textbf{Phase 3: Source Setup and Hybrid Assembly}
\begin{itemize}
\item Position Horn: Align the feed horn precisely at the center of the hemisphere.
\item Check Port: Verify the waveguide port orientation and matching.
\item Run Primary Sim: Simulate the horn and smooth-platform subsystem.
\item Use Hybrid Solver: Leverage the hybrid assembly solver for the horn antenna and lens including SiO$_2$.
\end{itemize}
\textbf{Phase 4: Field Export and Final Evaluation}

\section{Performance Evaluation and Discussion}
\begin{table}[!t]
\caption{Radiation performance for different beam-steering angles. These reported antenna efficiencies account for a 5\% degradation due to performance imperfections.}
\label{tab:radiation}
\centering
\footnotesize
\renewcommand{\arraystretch}{1.05}
\setlength{\tabcolsep}{4pt}
\begin{tabular}{c c c c}
\toprule
\textbf{$(\theta,\phi)$ (deg.)} &
\textbf{Directivity} &
\textbf{HPBW} &
\textbf{SLL} \\
&
\textbf{(dBi)} &
\textbf{(deg.)} &
\textbf{(dB)} \\
\midrule
$(0,0)$       & 14.72 & 21.6 & $-8.5$ \\
$(23,0)$      & 12.15 & 39.9 & $-9.4$ \\
$(22,150)$    & 11.96 & 39.6 & $-9.6$ \\
$(23,300)$    & 12.09 & 39.4 & $-9.8$ \\
$(50,0)$      & 13.29 & 22.7 & $-8.8$ \\
$(50,150)$    & 13.24 & 21.3 & $-7.7$ \\
$(76,360)$    & 11.38 & 41.8 & $-6.8$ \\
$(78,150)$    & 11.70 & 40.1 & $-9.2$ \\
$(78,300)$    & 11.43 & 37.4 & $-7.7$ \\
\bottomrule
\end{tabular}
\end{table}
\begin{table}[!t]
\centering
\caption{Antenna efficiency for different beam-steering directions.}
\label{tab:efficiency_scan}
\renewcommand{\arraystretch}{1.1}
\scriptsize
\begin{tabular}{ccc}
\toprule
\textbf{$\theta$ (deg.)} &
\textbf{$\phi$ (deg.)} &
\textbf{Antenna Efficiency (\%)} \\
\midrule
0  & 0   & 79 \\
23 & 300 & 78 \\
50 & 150 & 80 \\
78 & 150 & 71 \\
77 & 300 & 72 \\
\bottomrule
\end{tabular}
\end{table}
The discussion investigates how variations in the applied voltages across the 121 graphene sectors, Fig.~\ref{fig:graphene-sectors}, influence the performance of our proposed transmitarray antenna, Fig.~\ref{fig:transmitarray_antenna}.
It is evident from the radiation patterns in Fig.~\ref{fig:scanning_angles} that variations in the applied voltages across the 121 graphene sectors substantially change the main-lobe directions of our proposed transmitarray antenna.
Compared with the available literature, Tables~\ref{tab:beam_steering_comparison_1} and ~\ref{tab:beam_steering_comparison}, the proposed transmitarray antenna achieves substantially greater steering of the main radiation lobes. This behavior mainly results from the hemispherical lens geometry, which introduces additional spatial variations in the S$_{21}$ transmission phase, in conjunction with the bias voltages applied to the graphene sectors.
Furthermore, unlike planar reflectarrays, the proposed transmissive configuration benefits from the curved aperture geometry, which provides a broader phase compensation capability and enables more effective control of the outgoing wavefront. 
Integrating 121 independently tunable graphene sectors enhances phase resolution, allowing finer adjustment of the local transmission phase and improving the accuracy of beam pointing. 
Moreover, leveraging graphene’s tunable surface impedance and the multilayer electromagnetic interactions embedded in the S$_{21}$ transmission response extends the available phase-tuning range, enabling larger spatial phase gradients across the aperture and consequently increasing the achievable beam steering angles.

The results in Table~\ref{tab:radiation} demonstrate the capability of our proposed transmitarray antenna to achieve wide-angle two-dimensional beam steering while maintaining stable radiation characteristics.
Some SLL values reported Table~\ref{tab:radiation} that drop below $-8 ~\mathrm{dB}$ do not imply that the proposed transmitarray antenna is incapable of tracking moving targets. Nevertheless, the corresponding radiation patterns still exhibit a good agreement with the desired beam-steering behavior and can support moving-target tracking, but with lower probabilities or reduction in tracking moving targets.

The proposed antenna maintains high directivity across a wide scanning range, including extreme elevation angles. 
Specifically, at $\theta = 78^\circ$, it achieves directivities of 11.70 dBi for $\varphi = 150^\circ$ and 11.43 dBi for $\varphi = 300^\circ$. 
Despite steering the beam near the horizon, this represents only an approximately 3-dB reduction from the broadside case. 
This robust performance demonstrates the effectiveness of the hemispherical transmitarray configuration in preserving aperture efficiency at large scan angles.
Although the HPBW increases for large steering angles due to a reduction in effective aperture projection and an increase in phase deviations at extreme scan directions, the proposed antenna preserves stable sidelobe performance, with SLL values maintained between $-6.8$ and $-9.8$ dB.

Table~\ref{tab:efficiency_scan} reports the antenna efficiency of the proposed transmitarray antenna for different steering angles of the elevation angle and azimuth angle.
The proposed antenna maintains high efficiency even at extreme steering directions, yielding 71\% and 72\% at \((\theta, \phi) = (78^\circ, 150^\circ)\) and \((77^\circ, 300^\circ)\), respectively. 
This corresponds to a reduction of less than 8\% compared with the broadside case at \((\theta, \phi) = (0^\circ, 0^\circ)\), despite steering the beam toward near-horizon directions. These results confirm the ability of the hemispherical transmitarray antenna to maintain efficient power transmission over the wide scanning range.
Furthermore, this high efficiency stems from combining a low-loss multilayer transmission structure with tunable graphene-based unit cells.
Phase distortions induced by scan-angle variations can be significantly mitigated by leveraging the intrinsic spatial phase compensation of the curved aperture geometry, while the voltage-controlled graphene sectors enable precise control of the local transmission phase with minimal amplitude degradation.
In addition to the performance degradation caused by imperfections, it is worth emphasizing that the proposed model employs the hybrid solver and unidirectional coupling setup, under which the effects of $S_{11}$ can be effectively mitigated, resulting in an improvement in the overall antenna efficiency. 
Consequently, the proposed antenna achieves wide-angle beam steering while maintaining high aperture efficiency, which is critical for tracking moving targets. 

The impedance matching performance of the proposed transmitarray antenna for different beam-steering directions is illustrated in Fig.~\ref{fig:five_rows}.
The magnitude of S$_{11}$ at the feedhorn port after integration remains below $-10$ dB over the entire frequency range from 200 to 300 GHz for all typical steering angles, including broadside and extreme off-axis directions.
The superior impedance stability over the wide scanning range is mainly attributed to the FCC gold patches in our proposed antenna. The scaling factor of 0.8 between consecutive FCC rings introduces multiple resonant responses with gradually varying resonant frequencies across the aperture. The overlap of these closely spaced resonances creates a broadband impedance response, allowing the antenna to maintain effective electromagnetic coupling over the entire operating bandwidth. 
Consequently, variations in the beam-steering directions do not significantly degrade the impedance matching performance.
\section{Conclusion}
In conclusion, we successfully demonstrated the first comprehensive CST Studio Suite implementation of the stacked multilayer hemispherical graphene-based transmitarray antenna operating at 250 GHz.
We developed a fully automated CST macro to generate the multilayer hemispherical structure and its 121 independently tunable graphene sectors, establishing an efficient framework for modeling complex, curved, programmable metasurfaces.

The numerical results have verified that voltage-controlled graphene sectors enable the wide-angle 3D beam steering while ensuring stable radiation characteristics, high antenna efficiency, and broadband impedance matching.

Unlike existing beam-steering mechanisms in reflectarrays and antennas, Tables~\ref{tab:beam_steering_comparison_1} and ~\ref{tab:beam_steering_comparison}, our proposed hemispherical transmitarray achieves wider steering angles through a curved aperture, multilayer electromagnetic interactions, and fine phase control using graphene sectors.

Owing to the overlapping resonances of the FCC elements across adjacent rings and the stable electromagnetic response of the multilayer structure, our proposed transmitarray maintains broadband impedance matching, $|S_{11}| < -10$~dB with a relative bandwidth of 40\% from 200 to 300~GHz, across all beam-steering directions.

Compared with the available literature, Tables~\ref{tab:beam_steering_comparison_1} and ~\ref{tab:beam_steering_comparison}, numerical results of the radiation patterns have verified that our proposed transmitarray achieves wide-angle 3D beam steering through accurate voltage-controlled transmission-phase tuning of 121 independently programmable graphene sectors. 

Our proposed hemispherical transmitarray antenna maintains high efficiency over a wide beam-scanning range. Even at the maximum steering angles, the antenna achieves efficiencies of 71\% and 80\%, representing less than an 8\% drop from the broadside performance. This high performance stems from the low-loss multilayer structure, tunable graphene unit cells, and the intrinsic phase compensation of the hemispherical aperture. 
These features enable wide-angle beam steering with high aperture efficiency for THz beam- and target-tracking applications.

\ifCLASSOPTIONcaptionsoff
  \newpage
\fi
\bibliographystyle{IEEEtran}
\bibliography{references}
\end{document}